\documentclass[fleqn,usenatbib,onecolumn]{mnras}
\usepackage{newtxtext,newtxmath}
\usepackage[T1]{fontenc}
\usepackage{ae,aecompl}
\usepackage{cancel}
\usepackage{soul}
\usepackage{multirow}
\usepackage[symbol]{footmisc}
\usepackage{tabularray}

\usepackage{graphicx}	% Including figure files
\usepackage{amsmath,amsfonts}	% Advanced maths commands
\usepackage{mathrsfs}
\usepackage{breqn}
\usepackage{placeins}
\usepackage{import}
\usepackage{ulem}
\usepackage{mathtools}
\usepackage[svgnames]{xcolor}

\makeatletter
\newcommand{\setword}[2]{%
  \phantomsection
  \def\@currentlabel{\unexpanded{#1}}\label{#2}%
}
\makeatother

\renewcommand{\vec}[1]{\boldsymbol{#1}}

\newcommand{\DS}{\displaystyle}
\usepackage{graphicx}% Include figure files
\usepackage{dcolumn}% Align table columns on decimal point
\usepackage{bm}% bold math

\newcommand{\bfB}{{\bm B}}

\title[Cosmic ray acceleration due to small-scale MHD turbulence]{Acceleration of cosmic rays in presence of magnetohydrodynamic fluctuations
at small scales}

\author[Kundu, Singh, Vaidya]{
Sayan Kundu$^{\rm 1,2}$\thanks{E-mail: sayan.astronomy@gmail.com}, {Nishant K. Singh$^{\rm 3}$}, {Bhargav Vaidya$^{\rm 1}$} 
\\
\\
$^{\rm 1}$Discipline of Astronomy, Astrophysics and Space Engineering, Indian Institute of Technology, Indore, Madhya Pradesh, India - 452020\\
$^{\rm 2}$Department of Physics, University of Bath, Claverton Down BA2 7AY, UK\\
$^{\rm 3}$Inter-University Centre for Astronomy \& Astrophysics, Post Bag 4, Ganeshkhind, Pune, India - 411007
}

\date{Accepted XXX. Received YYY; in original form ZZZ}

\pubyear{2022}

\begin{document}
\label{firstpage}
\pagerange{\pageref{firstpage}--\pageref{lastpage}}
\maketitle

% Abstract of the paper
\begin{abstract}
This work investigates the evolution of the distribution of charged particles (cosmic rays) due to the mechanism of stochastic turbulent acceleration (STA) in presence of small-scale turbulence with a mean magnetic field. 
STA is usually modelled as a biased random walk process in the momentum space of the non-thermal particles. 
This results in an advection-diffusion type transport equation for the non-thermal particle distribution function.
{Under quasilinear approximation, and by assuming turbulent spectra with power being available only in the sub-gyroscale range}, we find that the Fokker-Planck diffusion coefficients $D_{\gamma\gamma}$ and $D_{\mu\mu}$ scale with the Lorentz factor $\gamma$ as: $D_{\gamma\gamma}\propto\gamma^{-2/3}$ and $D_{\mu\mu}\propto\gamma^{-8/3}$.
{We consider Alfv\`{e}n and fast waves in our calculations, and find a universal trend for the momentum diffusion coefficient irrespective of the properties of the small-scale turbulence. 
Such universality has already been reported regarding the spatial diffusion of the cosmic rays, and, here too, we observe a universality in the momentum diffusion coefficient.}
Furthermore, with the calculated transport coefficients, we numerically solve the advection-diffusion type transport equation for the non-thermal particles.
We demonstrate the interplay of various mircophysical processes such as STA, synchrotron loss and particle escape on the particle distribution by systematically varying the parameters of the problem.
{We observe that the effect of the small-scale turbulence is more impactful for the high energy protons as compared to the electrons and such turbulence is capable of sustaining the energy of the protons from catastrophic radiative loss processes.
Such a finding is novel and helps us to enhance our understanding about the hadronic emission processes that are typically considered as a competitor for the leptonic emission for certain astrophysical systems.}
\end{abstract}


%%%%%%%%%%%%%%%%%%%%%%%%%
%Introduction
%%%%%%%%%%%%%%%%%%%%%%%%%
\section{Introduction}
The transport of non-thermal charged particles dictates the emission properties of various highly energetic astrophysical sources. 
Usually, this transport phenomenon is mediated by a turbulent magnetic field, which subsequently leads the particles to exhibit diffusive behaviour in both space and energy domains. 
Such a diffusive behaviour in energy is a crucial component in accelerating the particles via the Fermi mechanism since the efficiency and the rate of particle acceleration directly depend on the scattering due to the random magnetic fields.
This turbulent acceleration is speculated to occur in different astrophysical sources with a diverse set of physical conditions, from the solar atmosphere \citep{petrosian_2004,selkowitz_2004,bian_2012} up to more exotic objects, e.g. blazars, gamma-ray bursts and other relativistic outflows \citep{bykov_1996,tramacere_2011,lemoine_2016,asano_2016,xu_2017,asano_2018,xu_2019}.
The magnetic turbulence also dictates the confinement of these charged particles in various astrophysical systems \citep{vukcevic_2007,shalchi_2009}.

An analytical quasilinear model of diffusion \citep{jokipii_1966,jokipii_1973} has often been used to estimate diffusion coefficients when the turbulent field is weaker than the background magnetic field.
Such analytical approach has been invoked to study the acceleration of particles via various MHD modes [by Alfv\'{e}n modes \citep{schlickeiser_1989,chandran_2000,cho_2006}; by compressive modes \citep{schlickeiser_1998,yan_2002,chandran_2003}].
For strong turbulence, on the other hand, several studies have employed numerical simulations to examine the transport properties of charged particles \citep{giacalone_1999,cesse_2001,candia_2004,fatuzzo_2010}. 
Most of these studies have focused their attention on the situation in which large-scale turbulence cascades toward smaller dissipative scales and the interaction between turbulent waves and charged particles is mediated by particular resonance conditions. 
Further, in such studies, it is also implicitly assumed that the particles’ gyro-radius is smaller than the maximum scale of the turbulence spectrum. 

{However, in certain astrophysical scenarios, the particle's gyro-radius can exceed the maximum coherence length of the turbulence. 
One example is the transport of supra-thermal particles near a relativistic shock, where sub-gyroscale turbulence is essential in scattering cosmic rays (CRs) and enabling them to complete enough Fermi cycles for efficient acceleration \citep{lemoine_2006b}. 
Additionally, such situations can also arise when the gyro-radius of highly energetic charged particles become comparable to albeit less than the Hillas limit of the system exceeding any length scale where fluctuations can occur in the system \citep{reichherzer_2022_proceeding}.}
Despite the wide application of this regime in several astrophysical systems, it has gained little attention to date. 
Although some recent research has been devoted to studying this under-explored field, it has primarily focused on the problem of spatial transport \citep{cesse_2001}. 
For example, \cite{plotnikov_2011} developed an analytical formulation of the spatial transport coefficients compatible with the numerical results for an intense small-scale random magnetic field. 
Furthermore, the work by \cite{subedi_2017} is worth noting, which studies {the spatial diffusion of the } charged particle in three-dimensional isotropic turbulent magnetic fields without a mean field.
{\cite{dundovic_2020} studied the transport of energetic particles in a synthetic magnetostatic turbulence, which in a way extended the work by \citep{subedi_2017}.}

In this {work}, we examine the momentum diffusive transport of charged particles in high-energy (or rigidity) regime with $R_{l}/l_{c}>>1$, where $R_{l}$ is the gyro-radius of the particle and $l_{c}$ is the highest correlation length of the turbulence.
{A possible scenario that illustrates this concept involves a particle undergoing acceleration through turbulent processes in a large-scale turbulent environment. 
As the particle continues to accelerate, it will eventually reach a point where its gyro-radius exceeds the correlation length of the turbulence that is accelerating it. 
This work seeks to address the question of whether the particle's motion will continue to be influenced by the turbulence, despite having surpassed its correlation length. 
To investigate this question, we focus on a turbulence spectrum that exhibits power at scales smaller than the gyro-radius of the particle, but not at the scale of the gyro-radius itself.}
% {We consider turbulence spectra which for the first time reveals the acceleration of the non-thermal particles in a situation where turbulence power is absent at the gyro-scale of the particle, but present in the sub-gyro regime. }
Using an asymptotic analysis of the quasilinear diffusion coefficient, we estimate the transport coefficients corresponding to this regime.
We also demonstrate the impact of the interplay between stochastic Fermi acceleration due to small-scale turbulence and synchrotron loss on the spectrum of non-thermal particles.

The paper is organised in the following way: in section~\ref{sec:calc} we show the calculations for the momentum transport coefficient for small-scale turbulence and the results are shown in sections~\ref{sec:dpp}, \ref{sec:dmumu} and \ref{sec:anisop}. 
In section~\ref{sec:trans_dist}, we show the results from solving the cosmic ray Fokker-Planck equation using the calculated transport coefficient.
{Subsequently in section~\ref{sec:application}, we discuss about possible astrophysical situations where the phenomena of small-scale turbulence can become potentially impactful on the energy distribution of the non-thermal particles. }
We discuss and summarize our findings in section~\ref{sec:diss} and in the appendix we lay-out all the required derivations.

%%%%%%%%%%%%%%%%%%%%%%%%%
%Calculation
%%%%%%%%%%%%%%%%%%%%%%%%%
\section{Calculation of the transport coefficients due to small-scale turbulence} \label{sec:calc} 

In this section, we provide the derivation of the momentum transport coefficients $D_{pp}$, $D_{\mu\mu}$ and $D_{\mu p}$ due to sub-gyroscale perturbations in the presence of a mean magnetic field, where $p$ and $\mu$ are the momentum and pitch-angle of the non-thermal particles, respectively.
We begin with the following form of the diffusive transport coefficients in momentum space \citep{schlickeiser_1993}:
\begin{eqnarray}
    \left[ \begin{array}{c}
    D_{\mu\mu}\\
    D_{\mu p}\\
    D_{pp}
    \end{array}\right] = \frac{\Omega ^{2}(1-\mu ^{2})}{2B_{0}^{2}}\left[ \begin{array}{c}
    1\\
    mc\\
    m^{2}c^{2}
    \end{array}\right] {\mathcal{R}}e \sum _{n=-\infty }^{n=\infty }
    \int_{k_{\rm min}}^{k_{\rm max}}\, d^{3}\vec{k}
    \int _{0}^{\infty }dte^{-\iota(k_{\parallel }v_{\parallel }-\omega +n\Omega )t} \left\{ J_{n+1}^{2}\left(\frac{k_{\perp }v_{\perp }}{\Omega}\right)\left[ \begin{array}{c}
    P_{{\cal RR}}({\vec{k}})\\
    T_{{\cal RR}}({\vec{k}})\\
    R_{{\cal RR}}({\vec{k}})
    \end{array}\right] \right. \nonumber \\
    + J_{n-1}^{2}\left(\frac{k_{\perp }v_{\perp }}{\Omega }\right) \left[ \begin{array}{c}
    P_{{\cal LL}}({\vec{k}})\\
    -T_{{\cal LL}}({\vec{k}})\\
    R_{{\cal LL}}({\vec{k}})
    \end{array} \right] + J_{n+1}\left(\frac{k_{\perp }v_{\perp }}{\Omega }\right)J_{n-1}\left(\frac{k_{\perp }v_{\perp }}{\Omega }\right)
    \left. \left[ e^{\iota 2\phi }\left[ \begin{array}{c}
    -P_{{\cal RL}}({\vec{k}})\\
    T_{{\cal RL}}({\vec{k}})\\
    R_{{\cal RL}}({\vec{k}})
    \end{array}\right] +e^{-\iota 2\phi }\left[ \begin{array}{c}
    -P_{{\cal LR}}({\vec{k}})\\
    -T_{{\cal LR}}({\vec{k}})\\
    R_{{\cal LR}}({\vec{k}})
    \end{array}\right] \right] \right\} \label{eq:QLTdiff} 
    \end{eqnarray}
where $\Omega = \Omega_{NR}/\gamma$ and $m=\gamma m_e$ are the gyro-frequency and mass of the relativistic non-thermal particle (electron for our case), respectively;
$\Omega_{NR}$ denotes the non-relativistic gyro-frequency and
$\gamma$ is the Lorentz factor;
wave number $k_{\rm min}$ corresponds to the inverse of some maximal length scale L as $k_{\rm min}=2\pi L^{-1}$,
and $k_{\rm max}$ corresponds to the dissipation scale; $v_{\perp }$ {and $k_{\perp}$ are} the particle's velocity and the wave vector components perpendicular to the mean magnetic field $\bfB_0 = B_0 \hat{z}$;
$\phi$ represents the phase angle between the Cartesian components of the wave vector
in a plane perpendicular to the mean magnetic field, i.e., $\phi =\tan^{-1} (k_{y}/k_{x})$;
${\cal L}$ and ${\cal R}$ represent left and right hand polarizations,
given by ${\cal L},{\cal R}=(x\pm \iota y)/\sqrt{2}$, with $x$ and $y$ being the Cartesian coordinates, and $\iota=\sqrt{-1}$ is the imaginary number;
$J_{n}(:)$ is the Bessel function of first kind with integer order $n$. 
Note that the {diffusive transport} coefficients in Eq.~(\ref{eq:QLTdiff}) corresponds to the lowest order in $V_{A}/c$ with $V_{A}$ being the Alfv\'en velocity. The above transport coefficients are only valid for Alfv\'en modes, whereas for other compressible modes, additional terms are needed to be considered in the equation for $D_{\mu p}$.
Here we focus only on Alfv\'{e}n waves because of their damping free nature in fully ionised medium \citep{kulsrud_1969,Ginzburg_1970,yan_2002}.
Further, the terms $P_{{ij}}$, $T_{{ij}}$, $Q_{ij}$ and $R_{{ij}}$ are defined using two point correlations at scales $\vec{k}$ and $\vec{k'}$ as follows, 
\begin{equation} \label{eq:corr}
\begin{aligned}
\left\langle B_i(\vec{k})B_j^{*}(\vec{k'})\right\rangle &=& \delta (\vec{k}-\vec{k'})P_{ij }(\vec{k}), \qquad\qquad
\left\langle B_i(\vec{k})E_j^{*}(\vec{k'})\right\rangle &=& \delta (\vec{k}-\vec{k'})T_{ij }(\vec{k}),  \\
\left\langle E_i(\vec{k})B_j^{*}(\vec{k'})\right\rangle &=& \delta (\vec{k}-\vec{k'})Q_{ij }(\vec{k}), \qquad\qquad
\left\langle E_i(\vec{k})E_j^{*}(\vec{k'})\right\rangle &=& \delta (\vec{k}-\vec{k'})R_{ij }(\vec{k}).
\end{aligned}
\end{equation}
where $B_{i}$ {and $E_{i}$ are} the magnetic {and electric} field fluctuations.
We also define terms $C_{{ij}}$ and $Y_{ij}$ that relate the magnetic field and velocity correlations as follows:
\begin{eqnarray}\label{eq:red_corr}
\left\langle u_i(\vec{k})B_j^{*}(\vec{k'})\right\rangle/V_{A}B_{0}=\delta (\vec{k}-\vec{k'})C_{ij }(\vec{k}), \nonumber \\
\left\langle u_i(\vec{k})u_j^{*}(\vec{k'})\right\rangle/V_{A}^{2}=\delta (\vec{k}-\vec{k'})Y_{ij }(\vec{k}),
\end{eqnarray}

Further considering MHD turbulence, the Ohm's Law implies $\mathbf{E}(\vec{k})=-\frac{\mathbf{u}(\vec{k})}{c}\times \mathbf{B}_{0}$ {with $\mathbf{u}(\vec{k})$ being the Fourier component of the velocity fluctuation of the underlying MHD flow.} 
Adopting this expression of electric field in Eq~(\ref{eq:corr}) and using the definitions provided in Eq.~(\ref{eq:red_corr})
we obtain, 
\begin{eqnarray}\label{eq:corr_mhd}
T_{ij }(\vec{k})=-\frac{B_{0}^{2}V_{A}}{c}\epsilon_{jpz}C_{ip}\left(\vec{k}\right), \nonumber \\
Q_{ij }(\vec{k})=-\frac{B_{0}^{2}V_{A}}{c} \epsilon_{imz}C_{mj}\left(\vec{k}\right), \nonumber \\
R_{ij }(\vec{k})=\frac{B_{0}^{2}V_{A}^{2}}{c^2}\left[\delta_{ij}Y_{pp}\left(\vec{k}\right)-Y_{ji}\left(\vec{k}\right)\right]
\end{eqnarray}
where, $z$ is the Cartesian coordinate along the mean magnetic field (see appendix~\ref{sec:appendix1} for detailed derivation). 
{To explore the effect of sub-gyroscale fluctuations threaded by a coherent magnetic field on the momentum transport of the charged particles, we consider both the isotropic and anisotropic turbulence spectra.
In the isotropic scenario we consider Alfv\`{e}n and fast wave turbulence, whereas for the anisotropic case we only consider Alfv\`{e}n wave turbulence. 
Such considerations are motivated by the fact that in wave-turbulence framework fast wave turbulence is known to follow an isotropic spectrum and the Alfv\`{e}n wave turbulence follows a highly anisotropic spectrum \citep{cho_2002,yan_2002}.
However in literature isotropic Alfv\`{e}nic turbulence is also considered \citep[see for example,][]{brose_2016}.
Moreover from solar wind data, the turbulence is found to become isotropic at and below electron gyro-scale range and the wave-wave interaction is found out to have resemblance with kinetic Alfv\'{e}n waves \citep{kiyani_2012}.
In the following sections first we calculate the momentum diffusion coefficients following an isotropic single scale injection spectrum and subsequently we carry out the calculation with a more realistic anisotropic turbulence. 
}

\subsection{Isotropic turbulence}\label{sec:iso_turb}
{In this section we focus on an isotropic single scale turbulence injection spectrum to compute the momentum transport coefficient for high-energy charged particles.
In particular we consider the following spectrum,}
\begin{eqnarray}\label{eq:turb_spec}
    Y_{ij}\left(\vec{k}\right) = \left(\delta_{ij}-\frac{k_{i}k_{j}}{k^2}\right)P_{iso}\delta\left(\frac{k}{m'k_{g}}-1\right)\,k^{-2}
\end{eqnarray}
where, $k_{g}$ is the inverse of the non-relativistic gyro-radius of a charged particle, $k_{g}=\Omega_{NR}/v$, and $m'$ is a parameter which dictates the scale of the turbulent energy injection. 
The choice of such a monochromatic injection spectrum is driven by the expectation that the energy present in the outer scale of the turbulence would maximally impact the high rigidity particles.  
Additionally, it has already been observed that the behavior of these particles is only marginally influenced by the specific form of the turbulence spectrum \citep[][]{subedi_2017}.
An estimation of $P_{iso}$ in the definition of turbulent spectrum can be computed following the equipartion {between the total} magnetic energy {and kinetic energy} \citep{yan_2002},
\begin{eqnarray}\label{eq:equipartition}
    \int d^{3}\vec{k} Y_{ii}\frac{\rho V_{A}^{2}}{2}\sim \frac{B_{0}^{2}}{8\pi} 
\end{eqnarray}
with $\rho$ being the density and $V_{A}=B_{0}^{2}/(4\pi\rho)$ is the Alfv\'{e}n velocity.
Comparing the value of the integration on the left side to the right side results in $P_{iso}\sim(8\pi m'k_{g})^{-1}$.
On substituting the correlation coefficients for MHD turbulence (Eqs.~\ref{eq:corr} and \ref{eq:corr_mhd}) in Eq.~(\ref{eq:QLTdiff}), we obtain the expression of $D_{pp}$ as follows (see appendix~\ref{sec:appendix2})
\begin{eqnarray}\label{eq:Dppinitial}
\nonumber
    D_{pp} = \frac{\Omega^2\left(1-\mu^2\right)}{2B_{0}^2}m^2c^2\frac{B_{0}^{2} V_{A}^{2}}{c^{2}}P_{0}\mathcal{R}e\left[\sum_{n = -\infty}^{\infty}\int_{k_{min}}^{k_{max}}\delta\left(\frac{k}{m'k_{g}}-1\right)k^{-2}d^{3}\vec{k} \int_{0}^{\infty}dt\,\,e^{-\iota\left(k_{||}v_{||}-\omega+n\Omega\right)t}
    \left\{J_{n+1}^2\left(\frac{k_{\perp}v_{\perp}}{\Omega}\right)\left(1+\frac{k_{\perp}^{2}}{2k^{2}}\right)\right .\right .\\
    \left .\left .\nonumber +J_{n-1}^2\left(\frac{k_{\perp}v_{\perp}}{\Omega}\right)\left(1+\frac{k_{\perp}^{2}}{2k^{2}}\right)+ 
    J_{n+1}\left(\frac{k_{\perp}v_{\perp}}{\Omega}\right)J_{n-1}\left(\frac{k_{\perp}v_{\perp}}{\Omega}\right)\left( e^{2\iota\psi}\frac{1}{2} \frac{k_{\perp}^{2}}{k^{2}} e^{-2\iota\psi} +e^{-2\iota\psi}\frac{1}{2} \frac{k_{\perp}^{2}}{k^{2}} e^{2\iota\psi}\right) \right\}\right] \\
    \nonumber
    = \frac{\Omega^2\left(1-\mu^2\right)}{2}m^2c^2\frac{V_{A}^{2}}{c^{2}}P_{0}\mathcal{R}e\left[\sum_{n = -\infty}^{\infty}\int_{k_{min}}^{k_{max}}\delta\left(\frac{k}{m'k_{g}}-1\right)k^{-2}d^{3}\vec{k} \int_{0}^{\infty}dt\,\,e^{-\iota\left(k_{||}v_{||}-\omega+n\Omega\right)t}
    \left\{\left(J_{n+1}^2\left(\frac{k_{\perp}v_{\perp}}{\Omega}\right)+J_{n-1}^2\left(\frac{k_{\perp}v_{\perp}}{\Omega}\right)\right)\left(1+\frac{k_{\perp}^{2}}{2k^{2}}\right) \right .\right.
    \\ \left .\left. 
    +J_{n+1}\left(\frac{k_{\perp}v_{\perp}}{\Omega}\right)J_{n-1}\left(\frac{k_{\perp}v_{\perp}}{\Omega}\right)\frac{k_{\perp}^{2}}{k^{2}} \right\}\right]
\end{eqnarray}
On performing the integration, $D_{pp}$ simplifies to (see appendix~\ref{sec:appendix3}),
\begin{eqnarray}\label{eq:Dppfinal}
\nonumber
    D_{pp} \simeq \Omega\left(1-\mu^2\right)m^2c^2\frac{V_{A}^{2}}{c^{2}}\pi^{2}{m'}k_{g} P_{0}\mathcal{R}e\left[ \int_{-1}^{1}dx
    \left\{\left(J_{\frac{\omega}{\Omega}-\frac{m'k_{g}xv\mu}{\Omega}+1}^2\left(\frac{m'k_{g}v}{\Omega}\sqrt{1-x^2}\sqrt{1-\mu^2}\right)\right.\right.\right. \\ \nonumber
    \left.\left.\left.
    +J_{\frac{\omega}{\Omega}-\frac{m'k_{g}xv\mu}{\Omega}-1}^2\left(\frac{m'k_{g}v}{\Omega}\sqrt{1-x^2}\sqrt{1-\mu^2}\right)\right)\left(\frac{3-x^{2}}{2}\right) \right.\right. \\ \left.\left.
    +\left(1-x^{2}\right)J_{\frac{\omega}{\Omega}-\frac{m'k_{g}xv\mu}{\Omega}+1}\left(\frac{m'k_{g}v}{\Omega}\sqrt{1-x^2}\sqrt{1-\mu^2}\right)J_{\frac{\omega}{\Omega}-\frac{m'k_{g}xv\mu}{\Omega}-1}\left(\frac{m'k_{g}v}{\Omega}\sqrt{1-x^2}\sqrt{1-\mu^2}\right) \right\}\right].
\end{eqnarray}
The above expression for $D_{pp}$ consists of several integrations involving Bessel function within the limits of $\pm 1$ and analytical solutions of such integrations are very challenging. 
We, therefore, consider integrating the above expression numerically to obtain the functional form of $D_{pp}$. 
{For that we further simplify Eq.(\ref{eq:Dppfinal}) by noting that the
Bessel function $J_{x}(y)$ contributes most significantly when $x\approx y$, i.e., when the order of the Bessel function is approximately equal to its argument. This gives,}
\begin{eqnarray}\label{eq:constarint_larger}
    \frac{\omega}{\Omega}-\frac{m'k_{g}xv\mu}{\Omega}\pm 1 \simeq \frac{m'k_{g}v}{\Omega}\sqrt{1-x^2}\sqrt{1-\mu^2} \gg 1 
\end{eqnarray}
which implies an analogous resonance condition of the following form: 
\begin{eqnarray}\label{eq:resonance_mod}
    \omega-m'\vec{k}_{g}\cdot \vec{v} \simeq \mp \Omega.
\end{eqnarray}
Furthermore, note that the presence of $\simeq$ in the above equation indicates that this condition has to be weakly satisfied.
Therefore, we introduce a parameter in order to modulate the value of $(\omega-m'\vec{k}_{g}\cdot \vec{v} \pm \Omega)$ {in order to broaden the resonance condition of Eq.~(\ref{eq:resonance_mod}).}
{Furthermore, it is important to highlight that the resonance condition mentioned in Eq.~(\ref{eq:resonance_mod}) is different from the quasilinear resonance expressed as,
$$k_{||}v_{||}-\omega\mp n\Omega=0.$$
Such a quasilinear resonance dictates the interaction between plasma waves and CR particles.
Whereas, the origin of Eq.~(\ref{eq:resonance_mod}) lies in the mathematical nature of the Bessel functions. 
In this context, we consider a broadening parameter that serves as a modulator for the difference between the left and right hand side of Eq.~(\ref{eq:resonance_mod}).
Hereafter, when referring to resonance broadening, we specifically refer to this type of broadening.
Additionally, it should be noted that the literature extensively discusses the broadening of the quasilinear resonance \citep[see][for example]{schlickeizer_2002,yan_2008}, but this work does not consider it.}
The presence of this resonance condition constrains the limit of the $x$ integral in Eq.~(\ref{eq:Dppfinal}). 
To identify the limits for Alfv\'{e}n waves, we undertake the following exercise: The resonance condition due to the shear Alfv\'{e}n wave, $(\omega = k_{||}V_{A} = kxV_{A})$, becomes,
\begin{eqnarray}
\label{eq:constraint}
\nonumber
\frac{\gamma m' k_{g} V_{A} x}{\Omega_{NR}} - \frac{m' k_{g} c \sqrt{1-\frac{1}{\gamma^2}}\mu x}{\Omega_{NR}}\gamma - \frac{m' k_{g}c \sqrt{1-\frac{1}{\gamma^2}}}{\Omega_{NR}}\gamma \sqrt{1-x^2}\sqrt{1-\mu^2} \simeq \mp 1 \\ 
\implies Ax-Bx-G\sqrt{1-x^2} = Q,
\end{eqnarray}
where the form of $A$, $B$ and $G$ are as follows,
$$A = \frac{\gamma m' \Omega_{NR} V_{A}}{\Omega_{NR}v} = \frac{\gamma \beta_{A} m'}{\sqrt{1-\frac{1}{\gamma^2}}};\quad B = \frac{m' \Omega_{NR} c\gamma \mu}{\Omega_{NR}v} \sqrt{1-\frac{1}{\gamma^2}} = \gamma \mu m';\quad G = \frac{m' \Omega_{NR} c\gamma}{\Omega_{NR}v} \sqrt{1-\frac{1}{\gamma^2}} \sqrt{1-\mu^2} = m' \gamma\sqrt{1-\mu^2}, $$
with $\beta_{A}$ being the Alfv\'{e}n velocity normalized to $c$, $\beta_{A}=V_{A}/c$.
Note that we have used the definition of $k_{g}$, while defining $A$, $B$ and $G$.
Further with the presence of $Q$, the resonance broadening effect can also be considered.
{Our interest is to find the range of $x$ such that the following equation is satisfied,
\begin{eqnarray}\label{eq:solutionx}
    Q_{min}\leq\frac{\gamma m' k_{g} V_{A} x}{\Omega_{NR}} - \frac{m' k_{g} c \sqrt{1-\frac{1}{\gamma^2}}\mu x}{\Omega_{NR}}\gamma - \frac{m' k_{g}c \sqrt{1-\frac{1}{\gamma^2}}}{\Omega_{NR}}\gamma \sqrt{1-x^2}\sqrt{1-\mu^2} \leq Q_{max}
\end{eqnarray}
}
Following the {range} of $x$ {through solving} Eq.~(\ref{eq:solutionx}), we write the form for $D_{pp}$ in the following way, 
{\begin{eqnarray}\label{eq:dpp}
\nonumber
    \frac{D_{pp}}{m_{e}^2c^2}=D_{\gamma\gamma} \simeq \Omega_{NR}\left(1-\mu^2\right)\gamma\beta_{A}^2\pi^{2}m'k_{g}P_{0}\mathcal{R}e\left[ \int_{\mathcal{F}^{+}}\,dy\,
    \left(\frac{3-y^{2}}{2}\right)J_{(A-B)y+1}^2\left(G\sqrt{1-y^2}\right)
    +\int_{\mathcal{F}^{-}}\,dy\,\left(\frac{3-y^{2}}{2}\right)J_{(A-B)y-1}^2\left(G\sqrt{1-y^2}\right) \right. \\ \left.
    +\int_{\mathcal{F}^{+}\cap\mathcal{F}^{-}}\,dy\,\left(1-y^{2}\right)J_{(A-B)y+1}\left(G\sqrt{1-y^2}\right)J_{(A-B)y-1}\left(G\sqrt{1-y^2}\right) \right].
\end{eqnarray}
where $\mathcal{F}^{+}$ corresponds to the range of $x$ when Eq.~(\ref{eq:solutionx}) is solved considering $Q_{max}=1+\sigma$ and $Q_{min}=1-\sigma$ with $\sigma$ being the broadening parameter.
Similarly $\mathcal{F}^{-}$ considers the range of $x$ for the solution of Eq.~(\ref{eq:solutionx}) with $Q_{max}=-1+\sigma$ and $Q_{min}=-1-\sigma$.}
{We solve Eq.~(\ref{eq:dpp}) for different values of $m',\, \beta_{A}\, \text{and}\, \sigma$ and the} result of the numerical integration is discussed in section~\ref{sec:dpp}.

Now we proceed to compute the form of $D_{\mu\mu}$
considering the correlation tensor $P_{ij}=B_{0}^{2}Y_{ij}$.
Such an assumption for the correlation function is typically used for Alfv\'{e}n waves \citep[see for example][]{yan_2002}.
With this correlation function and an exactly similar kind of calculation as shown in appendix~\ref{sec:appendix2} $\&$ \ref{sec:appendix3} leads to the following form for $D_{\mu\mu}$,
{\begin{eqnarray}\label{eq:dmumu}
\nonumber
    D_{\mu\mu} \simeq \frac{\Omega_{NR}}{\gamma}\left(1-\mu^2\right)\pi^{2}m'k_{g}P_{0}\mathcal{R}e\left[ \int_{\mathcal{F}^{+}}\,dy\,
    \left(\frac{1+y^{2}}{2}\right)J_{(A-B)y+1}^2\left(G\sqrt{1-y^2}\right)
    +\int_{\mathcal{F}^{-}}\,dy\,\left(\frac{1+y^{2}}{2}\right)J_{(A-B)y-1}^2\left(G\sqrt{1-y^2}\right) \right. \\ \left.
    +\int_{\mathcal{F}^{+}\cap\mathcal{F}^{-}}\,dy\,\left(1-y^{2}\right)J_{(A-B)y+1}\left(G\sqrt{1-y^2}\right)J_{(A-B)y-1}\left(G\sqrt{1-y^2}\right) \right].
\end{eqnarray}}
Here also, due to the presence of the Bessel function, we consider numerical integration. 
Interestingly, owing to the isotropic nature of the turbulence, all the components of the correlation function for $D_{\mu p}$ come out to be imaginary (see appendix~\ref{sec:appendix2}).
Therefore, $D_{\mu \gamma}=D_{\mu p}/(m_{e}c)$ does not make any contribution toward the transport of these non-thermal particles.

\subsection{Anisotropic turbulence}
In this section we calculate the momentum transport coefficient considering a realistic power law like turbulence spectrum for Alfv\`{e}n waves.
We further consider the maximum turbulence correlation length to be smaller than the gyro-radius of the charged particle by considering unit step function. 
In particular we choose the following form for the anisotropic turbulence spectrum,
\begin{eqnarray}\label{eq:aniso_spec}
    Y_{ij}(k)=P_{aniso}\left(\delta_{ij}-\frac{k_{i}k_{j}}{k_{\perp}^2}\right)\Theta\left(k_{\perp} - m'k_{g}\right)\delta\left(\frac{k_{||}}{m''k_{g}}-1\right)k_{\perp}^{-\alpha},
\end{eqnarray}
where $\Theta$ corresponds to Heaviside Theta function and $P_{aniso}$ being the injected turbulent power; $m'k_{g}$ and $m''k_{g}$ are the respective scales of $k_{\perp}$ and $k_{||}$ where the turbulence energy is being injected, with $k_{g}=\Omega_{NR}/v$; $k_{i}$ and $k_{j}$ corresponds to the components of wave vector $k$, in the perpendicular direction of the magnetic field.
The motivation behind choosing such a spectral form for the anisotropic turbulence spectrum stems from the observation that, at the largest length scale, MHD turbulence tends to exhibit a weak turbulent behaviour. 
In this regime, the energy cascade primarily occurs in the direction perpendicular to the mean magnetic field ($k_{\perp}$), while the parallel wavenumber ($k_{||}$) remains unchanged. 
In particular, the interaction between waves in weak turbulence leads to alterations in the perpendicular wavenumber while leaving the parallel wavenumber unaffected.
Therefore, considering that the maximum impact on high rigidity cosmic rays is influenced by the turbulence properties at the largest scale, we consider an anisotropic spectrum of the turbulence as Eq.~(\ref{eq:aniso_spec}).

With such a spectrum the equipartition of energy implies the form of $P_{aniso}$ as the following,
\begin{eqnarray}
    P_{aniso} \simeq\frac{\alpha-2}{2\pi m''k_{g}(m'k_{g})^{2-\alpha}}.
\end{eqnarray}
The positivity constraint of the power implies $\alpha>2$.
With such turbulence spectrum the momentum diffusion coefficients $D_{\gamma\gamma}$ becomes,
\begin{eqnarray}\label{eq:Dgammagamma_power_law}
    D_{\gamma\gamma}=\frac{D_{pp}}{m_{e}^2c^2} =\frac{\gamma\Omega_{NR}\left(1-\mu^2\right)}{4}2\pi^{2}\beta_{A}^{2} m''k_{g}P_{aniso}\mathcal{R}e\left[\int_{m'k_{g}}^{\infty}k_{\perp}^{-\alpha+1}d k_{\perp}\left\{J_{\frac{\omega}{\Omega}-\frac{m''k_{g}v_{||}}{\Omega}+1}^2\left(\frac{k_{\perp}v_{\perp}}{\Omega}\right)+J_{\frac{\omega}{\Omega}-\frac{m''k_{g}v_{||}}{\Omega}-1}^2\left(\frac{k_{\perp}v_{\perp}}{\Omega}\right)\right .\right .
    \\ \nonumber
    \left.\left. +2J_{\frac{\omega}{\Omega}-\frac{m''k_{g}v_{||}}{\Omega}+1}\left(\frac{k_{\perp}v_{\perp}}{\Omega}\right)J_{\frac{\omega}{\Omega}-\frac{m''k_{g}v_{||}}{\Omega}-1}\left(\frac{k_{\perp}v_{\perp}}{\Omega}\right) \right\} \right]
\end{eqnarray}
Although we have considered the upper limit of the integral to be $\infty$, note that the value of $k_{\perp}$ cannot take an arbitrarily large value owing to the following constraint,
\begin{eqnarray}\label{eq:aniso_const}
    m''k_{g}V_{A} - m''k_{g}v\mu\pm \Omega\approx k_{\perp}v\sqrt{1-\mu^{2}}
\end{eqnarray}
Following a similar kind of analysis described in the previous section, we introduce a parameter $Q$ and after some algebraic manipulations we write the above equation in the following way, 
\begin{eqnarray}\label{eq:mu_aniso}
    \frac{m''\beta_{A}}{\sqrt{1-\frac{1}{\gamma^{2}}}} - m''\mu+\frac{Q_{max}}{\gamma}\geq m'\sqrt{1-\mu^{2}}
\end{eqnarray}
Note that while writing the above equation we consider $k_{g}=\Omega_{NR}/\left(c\sqrt{1-1/\gamma^{2}}\right)$ and $k_{\perp}\geq m'k_{g}$.
From Eq.~(\ref{eq:mu_aniso}) we compute the range of $\mu$ such that the inequality is satisfied.
Subsequently for each value of $\mu$ in that range we compute the value of upper limit of the $k_{\perp}$ integral through the following equation,
\begin{eqnarray}
    k_{\perp}(Q)=\frac{\Omega_{NR}}{v\sqrt{1-\mu^{2}}\gamma}\left(\frac{\gamma m''\beta_{A}}{\sqrt{1-\frac{1}{\gamma^{2}}}}-\gamma m''\mu+Q\right).
\end{eqnarray}
$k_{\perp}(Q_{max})$ being the upper limit of the $k_{\perp}$ integral and the maximum between $k_{\perp}(Q_{min})$ and $m'k_{g}$ is considered as the lower limit of the integral. 
Following the limit of the integration we compute the integral in Eq.~(\ref{eq:Dgammagamma_power_law}) numerically.

%%%%%%%%%%%%%%%%%%%%%%%%%
%Results
%%%%%%%%%%%%%%%%%%%%%%%%%
\begin{figure}
    \centering
    \includegraphics[scale=0.31]{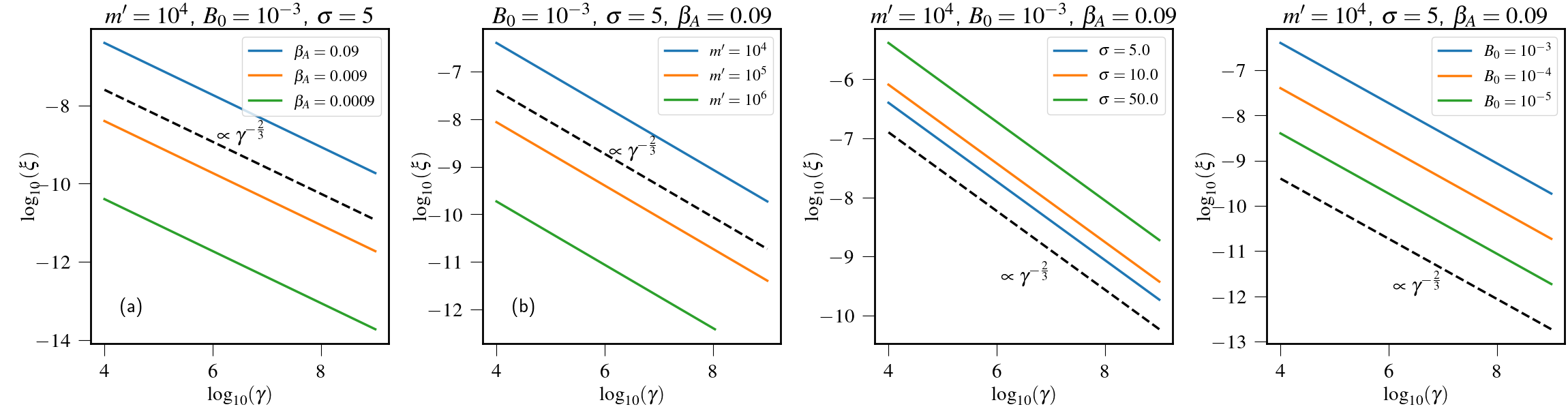}
    \caption{{Plot (isotropic case) showing the dependence of pitch-angle-averaged momentum diffusion coefficient on $\gamma$ for different values of Alfv\`{e}n velocity ($\beta_{A}$), turbulent injection scale ($m'$), broadening ($\sigma$) and magnetic field ($B_{0}$). All the plots show the same trend of $\xi\propto\gamma^{-{2}/{3}}$.}}
    \label{fig:diff_gamm}
\end{figure}

\section{Transport coefficients $D_{\gamma\gamma}$ and $D_{\mu\mu}$}
First, we present our results on the transport coefficients by numerically evaluating the integrals that appear in the expressions for $D_{\gamma\gamma}$ and $D_{\mu\mu}$ as presented above in section~\ref{sec:calc}. 
Subsequently, we solve the cosmic ray transport equation with the calculated diffusion coefficients in addition to the synchrotron loss process and study their interplay. 
\subsection{Momentum diffusion Coefficient $\left(D_{\gamma \gamma}\right)$}\label{sec:dpp}
We are interested here in the average diffusion which is obtained by integrating $D_{\gamma\gamma}$ over the distribution of the pitch angle $\mu\in[-1,1]$ {as the following,}
\begin{eqnarray}\label{eq:xi}
    \xi=\int_{-1}^{1}D_{\gamma\gamma}\,d\mu
\end{eqnarray}
In Fig.~\ref{fig:diff_gamm} we show the dependence of {pitch-angle-averaged momentum diffusion coefficient} {($\xi$)} on $\gamma$ for different values of the parameters $\beta_{A}$, $m'$, $\sigma$ and $B_{0}$, which are all defined above in section~\ref{sec:calc}.
For all the plots shown in the figure, $\xi$ exhibits a power-law trend following the same exponent with the Lorentz factor $\gamma$ of the non-thermal particles, $\xi \propto\gamma^{-2/3}$. 
In {panel (a)}, one can observe the increase in $\xi$ with increasing $\beta_{A}$ for a constant value of {$m'=10^{4}, B_{0}=10^{-3}\,\text{G}\,\text{and}\,\sigma=5$}.
This indicates that the higher the velocity of the Alfv\'{e}n wave is, the quicker the non-thermal particles will diffuse in $\gamma$ space.
In {panel (b)} we show the dependence of $\xi$ on $\gamma$ for different values of $m'$.
We observe that with increasing $m'$, $\xi$ decreases for a fixed $\gamma$ value.
Such behaviour of $\xi$ is expected as $m'$ parameterizes the scale of the turbulent energy injection, and higher values of $m'$ indicate that the energy is getting injected at a lower scale. 
The behaviour shown in {panel (b)} of the figure implies that such an energy injection at smaller scales reduces the momentum diffusion, resulting in the reduced acceleration of particles with large gyro-radii. 

In {panel (c)} {the right panel} of the figure we show the functional dependence of $\xi$ on $\gamma$ for fixed $\beta_{A}$, {$B_{0}$} and $m'$ but varying $\sigma$.
We observe with increasing $\sigma$ values the value of $\xi$ increases for a fixed $\gamma$ which is expected as with increasing $\sigma$, more and more Alfv\'{e}n waves would interact with the particles resulting in higher momentum diffusion $\xi$.
{Finally, in panel (d) we show the variation of the $\xi$ with the Lorentz factor $\gamma$ for different values of the magnetic field $B_{0}$.
From the trend it can be observed that with increasing $B_{0}$ value the diffusion coefficient increases which implies that with higher magnetic field the diffusion enhances.}

\subsection{Pitch-angle diffusion coefficient $\left(D_{\mu\mu}\right)$}\label{sec:dmumu}
\begin{figure}
    \centering
    \includegraphics[scale=0.31]{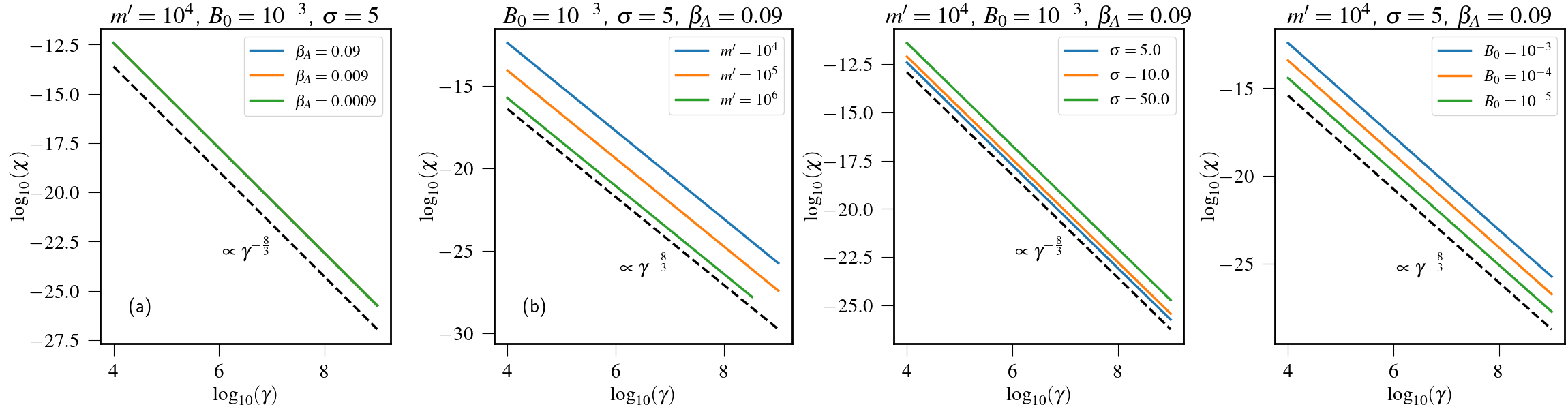}
    \caption{{Plot (isotropic case) showing the dependence of pitch-angle-averaged pitch angle diffusion coefficient on $\gamma$ for different values of Alfv\`{e}n velocity ($\beta_{A}$), Turbulence injection scale ($m'$), broadening ($\sigma$) and magnetic field ($B_{0}$). All the plots show the same trend of $\chi\propto\gamma^{-{8}/{3}}$.} %{$\int_{-1}^{1}D_{\mu\mu}\,d\mu\propto\gamma^{-8/3}$}, which is depicted by the black dashed line.
    }
    \label{fig:pitch_angle}
\end{figure}

\noindent
Similar to Eq.~(\ref{eq:xi}), we define another dimensionless pitch-angle averaged diffusion coefficient
$\chi$ as:
\begin{eqnarray}
    \chi=\int_{-1}^{1}D_{\mu \mu}\,d\mu\,,
\end{eqnarray}
with $D_{\mu \mu}$ being calculated by numerically integrating Eq.~(\ref{eq:dmumu}) for the region of $x$ satisfying Eq.~(\ref{eq:solutionx}).
In Fig.~\ref{fig:pitch_angle}, we plot $\chi$ for different values of $m'\,,\beta_{A},\, B_{0}\, \text{and}\, \sigma$

All the plots exhibit an inverse power-law trend with Lorentz factor $\gamma$ of the cosmic rays.
For all the cases in the figure, we find the same power-law index of $-8/3$.
In {panel (a)} of the figure, the plots are shown for constant {$m',\,B_{0}\,\text{and}\,\sigma$} but varying $\beta_{A}$.
The curves can be observed to almost overlap for different $\beta_{A}$, indicating that $\chi$ has a very weak dependency on the velocity of the Alfv\'{e}n waves.          
In {panel (b)}, the form of $\chi$ has been shown for different $m'$ values while $B_{0},\,\beta_{A}\,\text{and}\,\sigma$ are kept constant.
Similar to $\xi$, here also we observe decrease in $\chi$ with increasing $m'$.
In {panel (c)} of the figure we show the plots for different values of $\sigma$.
We find that $\chi$ increases with $\sigma$, which is expected as larger $\sigma$ consequently implies that more number of waves are interacting with the charged particles.
This results in more efficient diffusion.
{In the rightmost panel, we plot the variation of $\chi$ with the Lorentz factor $\gamma$ for different magnetic field values, $B_{0}$.
Similar to $\xi$, we can observe the increase in the diffusion coefficient with increasing magnetic field.}

{\subsection{Momentum diffusion coefficient due to anisotropic small-scale turbulence } \label{sec:anisop}
\begin{figure}
    \centering
    \includegraphics[scale=0.31]{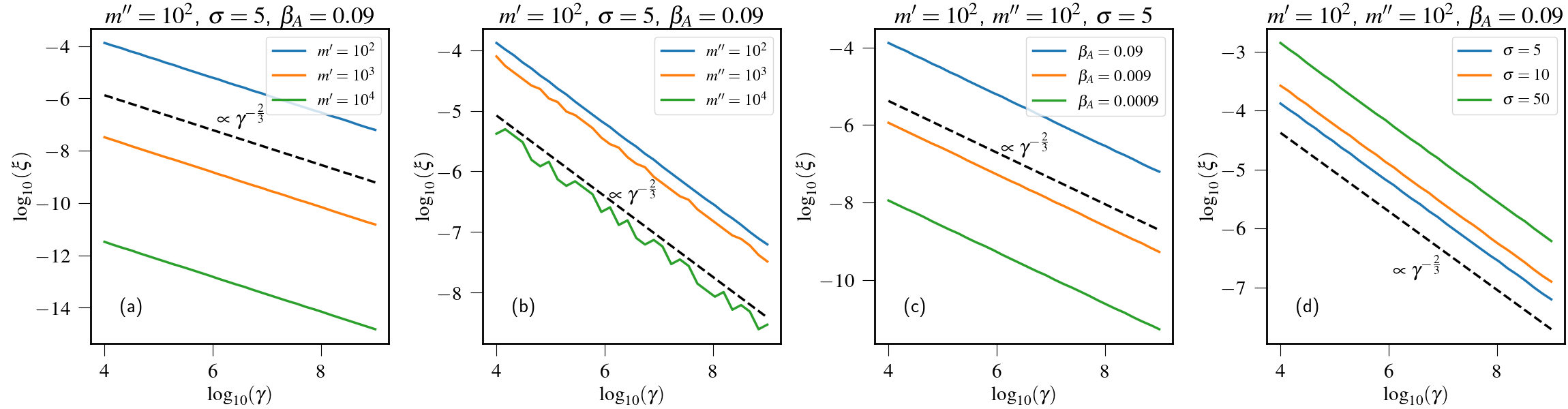}
    \caption{{Figure showcasing the dependence of the pitch-angle-averaged momentum diffusion coefficient ($\xi$) on particle Lorentz factor $\gamma$ for different parameter values when the underlying turbulence is anisotropic.
    Similar to the isotropic case, the diffusion coefficient can be observed to behave as a power-law with the particle Lorentz factor and with a similar index of $-2/3$.
    A black dashed curve of similar power-law trend is shown in all of the panel of the figure for the reference.}}
    \label{fig:anisotropic_dgammagamma}
\end{figure}
Fig.~\ref{fig:anisotropic_dgammagamma} presents the pitch-angle averaged momentum diffusion coefficient $\int_{-1}^{1}D_{\gamma\gamma}\,d\mu$ as a function of the particle Lorentz factor $\gamma$, for different parameter values. 
In all the simulations, the magnetic field value is fixed at $10^{-3}$\,G and we consider $\alpha=3$. 
The diffusion coefficient follows a power law behavior with an index of $-2/3$, which is consistent with the isotropic case.
Panel (a) shows the variation of the diffusion coefficient with different values of $m'$, which correspond to different scales of energy injection along the direction of $k_{\perp}$. 
It can be observed that for larger values of $m'$, the diffusion coefficient decreases. 
This trend is expected as a larger value of $m'$ corresponds to a smaller energy injection scale, resulting in a weaker effect of turbulence on the charged particles.
In panel (b), we modulate the value of $m''$ and observe its effect on the diffusion coefficient.
Similar to panel (a), the diffusion coefficient exhibits a decreasing trend with an increasing $m''$ value.
Additionally, as we increase the value of $m''$, we notice that the diffusion coefficient becomes highly responsive, resulting in a fluctuating pattern. Nevertheless, the general tendency is apparent, and it conforms to a power-law distribution with an exponent of $-2/3$. This behavior indicates that the injection of turbulence power at the coherent magnetic field's length scale has a more significant qualitative impact than the length scale perpendicular to $B_{0}$.
Next, in panel (c), we modulate the Alfv\`{e}n velocity of the small-scale Alfv\`{e}n waves and show the trend of the diffusion coefficient. 
It is observed that with decreasing Alfv\`{e}n speed of the underlying fluctuations, the momentum diffusion decreases.
Finally, in panel (d), we investigate the effect of the parameter $\sigma$ on the momentum diffusion coefficient. As expected, with an increasing $\sigma$ value, the momentum diffusion coefficient increases. 
This trend is due to the fact that particles interact with more waves as the value of $\sigma$ increases.}

{In summary, we have observed that the pitch-angle averaged momentum diffusion coefficient exhibits a power-law like behaviour with an index of $-2/3$ for both isotropic and anisotropic turbulence spectrum. 
This is consistent with theoretical expectations, as cosmic ray particles with high rigidity are expected to be weakly dependent on the specific form of the turbulence spectrum.}
\vskip4ex
{Note that, during all the above calculations of the diffusion coefficients, it was ensured that the values of both the order and argument of the Bessel functions remained sufficiently large to satisfy Eq.~(\ref{eq:constarint_larger}). 
In the majority of cases, the values of the argument and order were observed to be greater than 100, while in one case, it remained greater than 70. 
Moreover, our analysis indicates that the parameters $m'$ and $m''$ have a significant impact on modulating the values of the argument and order. 
Increasing their values leads to larger values of the argument and order. 
This behavior is expected, as $m'$ and $m''$ determine the scale difference between the gyro-radius and turbulence injection scale. 
A decrease in their values implies a reduction in the rigidity of the non-thermal particle and eventually leading to the case of large-scale turbulence. 
Hence, for large-scale turbulence, the order of the Bessel function is typically considered in the range of $0,\,\pm1,\,\pm2$ \citep{berezinskii_1990}.}

{Furthermore, we anticipate that for anisotropic turbulence, the pitch angle diffusion coefficient will follow a similar trend to that of the isotropic case (see Fig.~\ref{fig:pitch_angle}). 
This expectation is based on the comparable behavioral patterns displayed by the momentum diffusion coefficient for both isotropic and anisotropic turbulence. 
Additionally, in quasilinear theory, the momentum diffusion coefficient is related to the spatial diffusion coefficient along the direction of the coherent magnetic field through the constraint given by \citep{thornbury_2014}:
\begin{eqnarray}\label{eq:QLT}
\mathcal{K}D_{\gamma\gamma}=\frac{1}{9}\gamma^{2}V_{A}^{2},
\end{eqnarray}
where $\mathcal{K}$ represents the spatial diffusion coefficient along the direction of the coherent magnetic field $B_{0}$. 
The spatial diffusion coefficient is related to the pitch-angle diffusion coefficient \citep{shalchi_2009_book}. 
Thus, once the behavior of the momentum diffusion coefficient is known, the behavior of the pitch angle diffusion coefficient can be constrained by the above equation. 
Therefore, we abstain from explicitly calculating the behavioral trend of the pitch-angle-averaged pitch angle diffusion coefficient due to the anisotropic turbulence spectrum.}

\subsection{Solutions of the Fokker-Planck equation}\label{sec:trans_dist}
\begin{figure}
    \centering
    \includegraphics[scale=0.51]{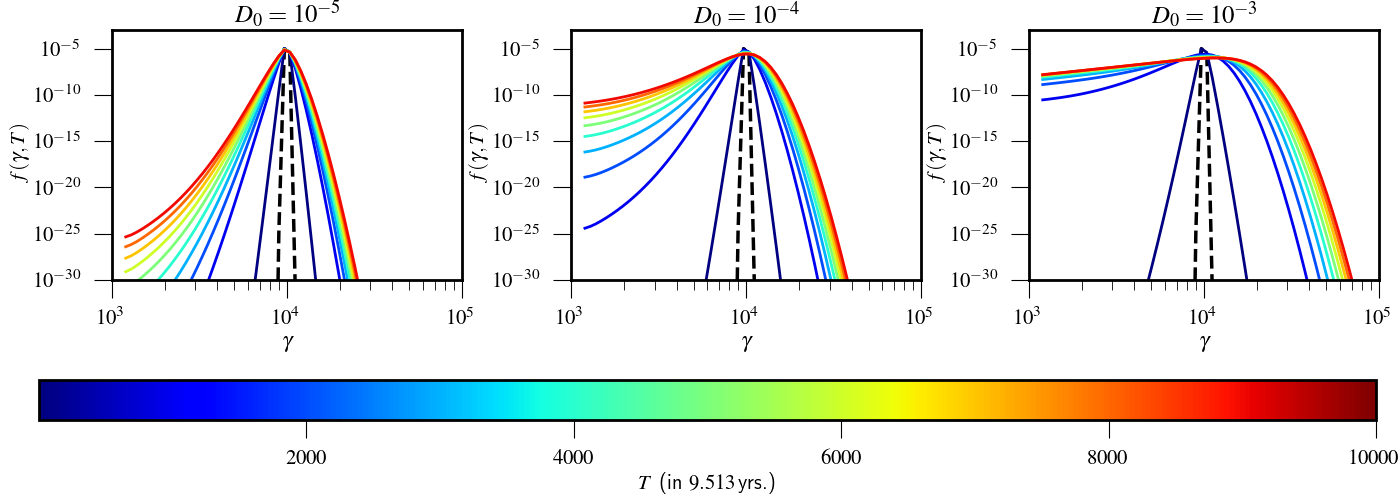}
    \caption{Evolution of an initial Gaussian (with mean $10^{4}$ and standard deviation $10^2$) for stochastic acceleration due to small-scale turbulence with different values of $D_{0}$ and following Eq.~(\ref{eq:acceleration}).
    The initial function is shown with a black dashed curve.}
    \label{fig:acceleration}
\end{figure}

In this section we demonstrate the effect of the small-scale turbulence on the non-thermal particle spectrum by numerically solving the cosmic ray transport equation with the coefficients calculated in the earlier sections.
Note that all the numerical simulations are performed with a conservative, second order accurate IMEX scheme \citep{Kundu_2021} and considering a discretization of the the particle Lorentz factor $\gamma$ from $\gamma_{\min}=10^{3}$ to $\gamma_{\max}=10^{7}$ with $128$ logarithmically spaced bins to provide equal resolution per decade.

In a turbulent medium with a guided field, the transport of cosmic rays is governed by a Fokker-Planck equation of the following type \citep[][]{kirk_1988,schlickeiser_1998},
\begin{eqnarray}\label{eq:schlickeizer}
    \frac{\partial F}{\partial t} =\frac{\partial}{\partial z}\left(\mathcal{K}\frac{\partial F}{\partial z}\right)-\left(U+\frac{1}{4p^{2}}\left(\frac{\partial}{\partial p}p^{2}va_{1}\right)\right)\frac{\partial F}{\partial z} + \left(\frac{p}{3}\frac{\partial U}{\partial z}+\frac{v}{4}\frac{\partial a_{1}}{\partial z}\right)\frac{\partial F}{\partial p}+\frac{1}{p^{2}}\frac{\partial}{\partial p}\left(p^{2}a_{2}\frac{\partial F}{\partial p}\right)+S_{0}
\end{eqnarray} 
where $F$ is the pitch angle averaged cosmic ray distribution function, $U$ is the non-relativistic fluid velocity, $v$ is the velocity of the cosmic ray particle and $p$ is the momentum of the cosmic rays and  
\begin{eqnarray}\label{eq:kas}
    \mathcal{K}=\frac{v^{2}}{8}\int^{1}_{-1}d\mu\,\frac{(1-\mu^{2})^{2}}{D_{\mu\mu}},\quad a_{1}=\int_{-1}^{1}d\mu\,(1-\mu^{2})\frac{D_{\mu p}}{D_{\mu\mu}},\quad  a_{2}=\frac{1}{2}\int_{-1}^{1}d\mu\,\left(D_{pp}-\frac{D_{\mu p}^{2}}{D_{\mu\mu}}\right).
\end{eqnarray}
Note that while deriving Eq.~(\ref{eq:schlickeizer}) the background flow and the guided magnetic field are considered to be in the same spatial direction, $z$, and the timescale of pitch-angle scattering is assumed to be minimum among all the timescales present in the system. 
The latter assumption introduces the spatial diffusion term parallel to the guided magnetic field in the right hand side of the equation \citep{shalchi_2020}.
A term consisting diffusion of cosmic rays in the direction perpendicular to the guided field also arises in the Fokker-Planck equation due to the stochasticity in the magnetic field line structure \citep{shalchi_2021}.
For the current work such term due to perpendicular diffusion is neglected as quasilinear theory is unable to address such diffusive transport \citep{shalchi_2020}. 
As noted in the end of section~\ref{sec:calc}, $D_{\mu p}=0$ which gives $a_{1}=0$ and $a_{2}$ is simply the pitch-angle averaged $D_{pp}$.
For this work we consider averaging out the spatial coordinates and following the leaky-box approximation \citep{lerche_1985} we replace the spatial diffusion and convection term by a momentum dependent escape term \citep{rieger_2019} with an escape timescale defined as $T_{\rm esc}\sim\mathcal{K}^{-1}\propto \gamma^{-8/3}$.
Further considering the calculated forms for pitch-angle averaged diffusion coefficients (see section~\ref{sec:dmumu}) in addition to synchrotron cooling and neglecting adiabatic loss, Eq.~(\ref{eq:schlickeizer}) takes the following form (see appendix \ref{sec:transport} for a derivation),
\begin{eqnarray}\label{eq:transport}
    \frac{\partial f}{\partial T} +\frac{\partial}{\partial \gamma}\left(2a\gamma^{-\frac{5}{3}}-\gamma^{2}\right)f=\frac{\partial}{\partial \gamma}\left(a\gamma^{-\frac{2}{3}}\frac{\partial f}{\partial \gamma}\right)-b\gamma^{\frac{8}{3}}f
    + S,
\end{eqnarray}
where $a$, $b$ and $S$ are defined via Eq.~(\ref{eq:coeff_mod}).

Before considering the interplay of various micro-physical processes, we first analyze the effect of the acceleration due to small-scale turbulence only. 
For that, we numerically solve the following equation which is similar to Eq.~(\ref{eq:transport}) but without the synchrotron loss, particle escape, and injection,
\begin{eqnarray}\label{eq:acceleration}
    \frac{\partial f}{\partial T} +\frac{\partial}{\partial \gamma}\left(2D_{0}\gamma^{-\frac{5}{3}}\right)f=\frac{\partial}{\partial \gamma}\left(D_{0}\gamma^{-\frac{2}{3}}\frac{\partial f}{\partial \gamma}\right),
\end{eqnarray}
with $D_{0}$ being a parameter with which the efficiency of acceleration can be tuned {, $\int_{-1}^{1}D_{\gamma\gamma}\,d\mu = D_{0}\gamma^{-2/3}$.
The value of $D_{0}$ can be determined from the dependence of the momentum diffusion coefficients on $\gamma$ as shown in Figs.~\ref{fig:diff_gamm}, \ref{fig:anisotropic_dgammagamma}, and observed to vary between $10^{-2}-10^{-5}$ for various parameter values typically observed in astrophysical systems.}

The numerical solution of Eq.~(\ref{eq:acceleration}) for different times and different $D_{0}$ values are shown in Fig.~\ref{fig:acceleration}.
An arbitrary Gaussian with $10^{4}$ and $10^{2}$ as mean and standard deviation respectively, has been considered as the initial distribution (shown by a black dashed curve). 
{Owing to the lower efficiency of the turbulent acceleration process due to small-scale fluctuations, we choose a larger final time ($\sim 100$\,kyr) to demonstrate sufficient acceleration of the initial Gaussian profile.}
With time all the plots show the spreading of the initial distribution owing to the momentum diffusion and acceleration due to small-scale turbulence. 
However, the spreading of the initial distribution function is not uniform, the low energy part spreads faster than the high energy.
Such an acceleration can be analyzed by observing the dependency of the acceleration timescale, $\tau_{acc}$, on $\gamma$, which is $\tau_{acc}\sim \gamma^2/D\propto\gamma^{8/3}/D_{0}$ from Eq.~(\ref{eq:acceleration}).
It clearly shows that the timescale of acceleration is smaller for smaller $\gamma$, which explains the faster acceleration in the low energy part. 
The acceleration timescale also inversely depends on the choice of $D_{0}$, which is why we observe faster acceleration for higher values of $D_{0}$.

After analyzing various aspects and features of stochastic acceleration due to small-scale turbulence we now proceed to analyze the interplay of different micro-physical processes with this acceleration. 
To explore the combined effect of different processes on the distribution function, we numerically solve Eq.~(\ref{eq:transport}) with the above-mentioned algorithm by incorporating different processes gradually.

\begin{figure}
    \centering
    \includegraphics[scale=0.51]{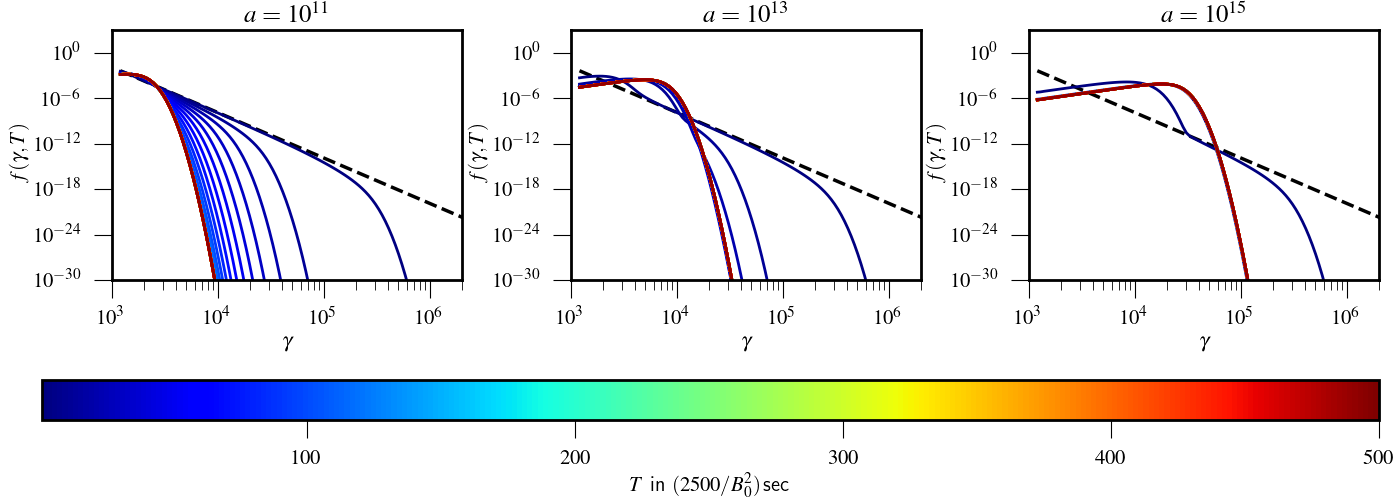}
    \caption{Evolution of an initial power-law energy distribution of the form $\gamma^{-6}$ following Eq.~(\ref{eq:transport}) considering synchrotron loss process and different values for $a$ for electrons.
    The values for $b$ and $S$ are considered zero.
    The initial distribution is shown with the black dashed curve.
    {Different color of the distribution function corresponds to different time of evolution, as illustrated by the colorbar. 
    To account for the varying magnetic field values observed in different astrophysical systems and the resulting variation in temporal units, the unit time is specified in terms of a variable magnetic field.
    }}
    \label{fig:different_a}
\end{figure}

In Fig.~\ref{fig:different_a} we show the effect of the interplay of synchrotron loss process and particle acceleration due to small-scale turbulence on the particle distribution function.
We solve Eq.~(\ref{eq:transport}) with an initial power-law type particle distribution of the form,  $f(\gamma,0)\propto\gamma^{-6}$.
As has been shown in the previous section~\ref{sec:dpp}, the transport coefficients are dependent on the choice of the broadening $\sigma$ and injected power $P_{0}$,
which could be different for different astrophysical systems.
As an illustration, therefore, we solve Eq.~(\ref{eq:transport}) to demonstrate the
effects induced by the interplay of various micro-physical processes on the distribution function by varying $a$ and $b$ which are treated as arbitrary parameters in this work.
Nonetheless, we make some estimates for the parameter $a$ {considering some generic values for the magnetic field and Alfv\`{e}n velocity in different astrophysical situations.}
Noting $a=D_{0} \gamma_s^{-11/3}/(c_0 B^2)$, {in Table~\ref{tab:quant} we give some quantitative estimate for the values of both $D_{0}$ and $a$ for different astrophysical environments. 
The quantitative estimations are shown for both electrons and protons.
{Note that} all estimates of the momentum diffusion coefficient presented in the table are computed from the isotropic turbulence case due to fewer parameter specifications.
It is important to emphasize that the quantitative presentation of diffusion coefficients for various astrophysical systems aims to provide an estimate and more importantly demonstrate the variation in diffusion values between electrons and protons.
However, it is crucial to acknowledge that the specific parameter values chosen for the calculations can influence the resulting diffusion coefficients.
If alternative parameter values were selected, the diffusion coefficients would differ, while the qualitative concept and trends would remain unchanged.

The value of $c_{0}$ is calculated to be $1.2\times 10^{-9}$ for electron and $2.08\times 10^{-19}$ for proton and also the value $m'$ for all the calculations is considered to be fixed at $10^{5}$.
Moreover, it can be noticed that the length scale for turbulence injection is not in the same order of scales typically where the injection of turbulence happens in those astrophysical systems.
Additionally one can observe that the values of the momentum diffusion coefficient is smaller for proton compared to that of the electron of same Lorentz factor. 
Such kind of difference in the momentum diffusion value results in longer acceleration time for the former as compared to the latter one. 
Further implication of such behaviour is explored in the following part of this section. }

\begin{table}
\centering
\label{tab:quant}
\begin{tblr}{
  cells = {c},
  cell{2}{1} = {r=2}{},
  cell{2}{3} = {r=2}{},
  cell{2}{4} = {r=2}{},
  cell{4}{1} = {r=2}{},
  cell{4}{3} = {r=2}{},
  cell{4}{4} = {r=2}{},
  cell{6}{1} = {r=2}{},
  cell{6}{3} = {r=2}{},
  cell{6}{4} = {r=2}{},
  vlines,
  hline{1-2,4,6,8} = {-}{},
  hline{3,5,7} = {2,5-7}{},
}
{\\ Environment}                       & { Particle \\Nature} & {Magnetic \\field \\ ($B_{0}$)} & {\\$\beta_{A}$}        & {\\$D_{0}$}              & {\\ $a=\frac{D_{0}}{c_{0}B_{0}^{2}}$} & {Relativistic\\gyro-radius\\($k_{g}^{-1}\,$cm)} \\
{Galaxy\\Cluster}                 & Electron            & $10\,\mu$G          & {$9\times 10^{-4}$\\ \citep{petrosian_2001}}  & $1.12\times 10^{-5}$ & $9.3\times 10^{13}$              & $1.7\times 10^{8}\,\gamma$                      \\
                                  & Proton              &                    &                    & $4.29\times 10^{-7}$ & $2.06\times 10^{22}$             & $3.13\times 10^{11}\,\gamma$                    \\
{Relativistic \\shock downstream} & Electron            & {$1\,$mG \\ \citep{virtanen_2005}}               & $9\times 10^{-2}$  & $4.54\times 10^{-3}$ & $3.78\times 10^{12}$             & $1.7\times 10^{6}\,\gamma$                      \\
                                  & Proton              &                    &                    & $1.73\times 10^{-4}$ & $8.32\times 10^{20}$             & $3.13\times 10^{9}\,\gamma$                     \\
{Interstellar\\Medium}            & Electron            & {$3\,\mu$G\\ \citep{farmer_2004}}          & {$6.6\times10^{-4}$\\ \citep{farmer_2004}} & $5.09\times 10^{-6}$ & $4.7\times 10^{14}$              & $5.69\times 10^{8}\,\gamma$                     \\
                                  & Proton              &                    &                    & $1.94\times 10^{-7}$ & $1.04\times 10^{23}$             & $1.04\times 10^{12}\,\gamma$                    
\end{tblr}
\caption{Quantitative estimate for the values of $D_{0}$ and $a$ for different astrophysical systems. Column~1 depicts the name of the astrophysical system, column~2 represents the nature of particle for which the values of $D_{0}$ and $a$ are calculated. Columns~3 and 4 represent typical values for the magnetic field and Alfv\`{e}n velocity in such astrophysical environments.  Column~5, 6 and 7 shows the numerical values for $D_{0}$, $a$ and relativistic gyro-radius considered for the calculations.}
\end{table}

\begin{figure}
    \centering
    \includegraphics[scale=0.51]{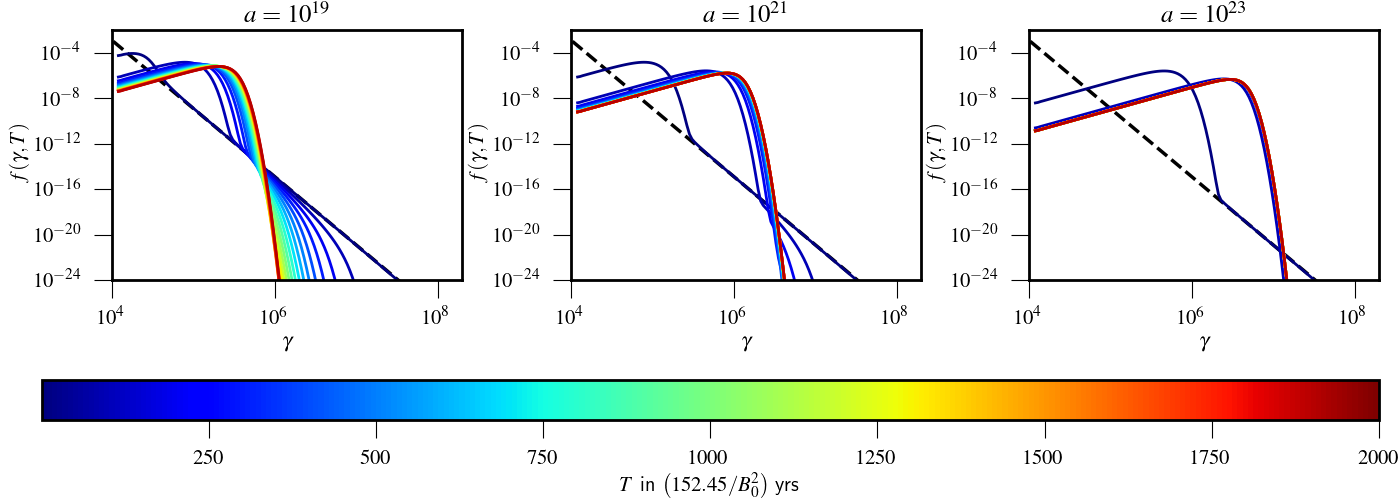}
    \caption{{Evolution of an initial power-law energy distribution of the form $\gamma^{-6}$ following Eq.~(\ref{eq:transport}) considering synchrotron loss process and different values for $a$ which typically occurs for protons.
    The values for $b$ and $S$ are considered zero.
    Different color of the distribution function corresponds to different time of evolution, as illustrated by the colorbar. 
    To account for the varying magnetic field values observed in different astrophysical systems and the resulting variation in temporal units, the unit time is specified in terms of a variable magnetic field.}}
    \label{fig:no_escape_proton}
\end{figure}

In the left panel of Fig.~\ref{fig:different_a} we show the evolution of the {energy} distribution function {for electrons} without the source and escape terms (considering $b=0$ and $S=0$ in Eq.~\ref{eq:transport}) and considering $a=10^{11}$.
{A color-coded representation is utilized to illustrate the evolutionary trend of the distribution function over time. 
Different colors correspond to different points in time, as shown on the color bar located at the bottom of the figure. 
The unit time is specified based on a variable magnetic field ($B_{0}$). 
Such a choice of temporal unit is motivated by the fact that the synchrotron cooling time varies in different astrophysical environments due to differences in the magnetic field values. 
Therefore use of such temporal unit allows for the consideration of different synchrotron cooling times, depending on the magnetic field values present in different astrophysical systems.
The use of this representation allows for a better understanding of the evolution of the distribution function in different astrophysical systems. 
To aid in the comprehension of the results, the unified temporal unit is used in all subsequent figures.}
The distribution function {can be observed to develop} an exponential cut-off at higher $\gamma$, which moves towards lower energy as time progresses, due to synchrotron cooling.
Additionally a hump like structure can be observed to develop at the low energy regime due to the acceleration of low energy particles owing to the turbulent acceleration.
The overall distribution function attains a steady state as a result of the competition between stochastic acceleration and synchrotron loss. 
The form of the steady state distribution can be computed analytically from Eq.~(\ref{eq:transport}) considering $b=0$ and $S=0$ and is $\propto \gamma^{2}\,\exp\left\{-\Lambda(\gamma)\right\}$, where $\Lambda(\gamma)=3\left(\gamma^{{11}/{3}} - 1\right)/11a$ \citep[see Eq.~(A1) of][]{Kundu_2021}. 
Further, with such a steady state distribution function it can be observed that the maximum of the distribution occurs at $\left(2a\right)^{3/11}$ and it increases with $a$ which can be observed from the middle and the right panel of the figure where the evolution of the distribution is shown form $a=10^{13}$ and $a=10^{15}$ respectively.

{Fig.~\ref{fig:no_escape_proton} illustrates the evolution of the distribution function for protons.
All the panels of the figure show that the steady state distribution function for protons is morphologically very similar to that of electrons, although with variations in the peak positions.
However, the time taken for protons to attain the steady state distribution function is longer than that for electrons as can be observed comparing the temporal unit and final time of both the figures. 
This difference is attributed to the fact that the synchrotron cooling time for protons is higher due to their higher rest mass, and the momentum diffusion coefficient for protons is lower than that for electrons of similar energy, which can be inferred from the Column~5 of Table~\ref{tab:quant}. 
This implies that it takes more time to accelerate a proton compared to an electron of the same Lorentz factor.
As a result, the value of the parameter $a$ is larger for protons, leading to a longer time to reach the steady state.}

{Further observation reveals that the value of $\gamma$ at which the distribution function is maximum for protons is higher than that for electrons. 
This indicates that small-scale MHD fluctuations can sustain the energy of higher energy protons for a longer period of time than electrons from the catastrophic synchrotron cooling. 
Therefore, it can be concluded that the effect of small-scale turbulence would be more prominent for higher energy protons than for electrons.}

{It is important to note that the evolution of the distribution function show a similar trend for both electrons and protons, and the steady state distribution functions are morphologically similar. 
Therefore, we focus on the electron distribution for all the subsequent analyses, but the results can be extended to the proton distribution in a similar manner.
}

\begin{figure}
    \centering
    \includegraphics[scale=0.5]{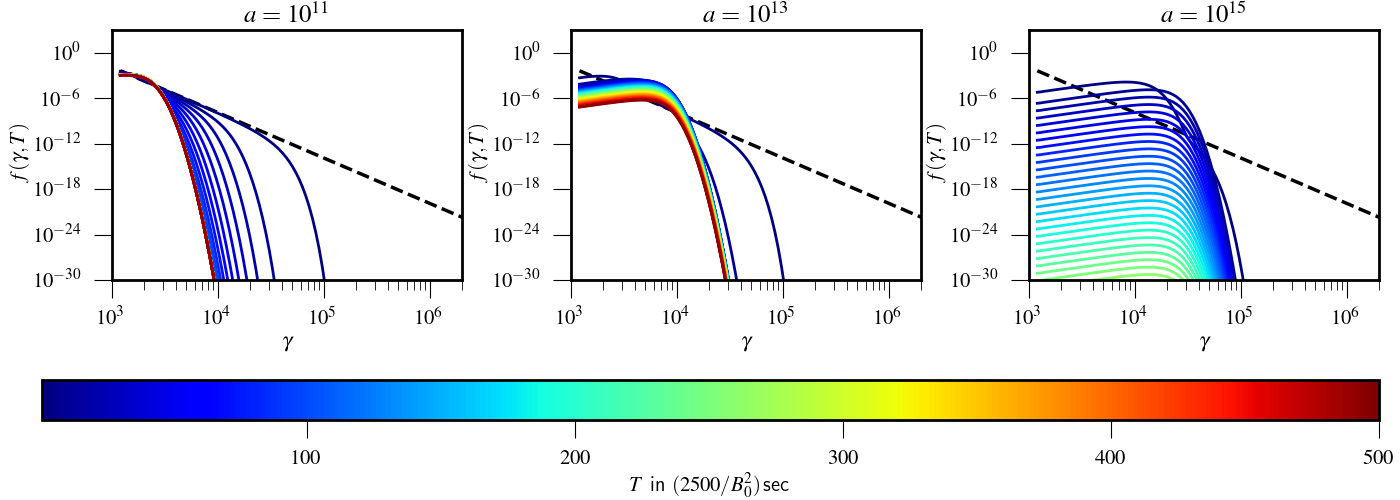}
    \caption{Evolution of an initial power-law energy distribution of the form $\gamma^{-6}$ following Eq.~(\ref{eq:transport}) considering synchrotron loss process with different values for $a$ and $b=10^{-6}$.
    The values for $S$ is considered as zero.
    The initial distribution is shown with the black dashed curve.
    {Different color of the distribution function corresponds to different time of evolution, as illustrated by the colorbar. 
    To account for the varying magnetic field values observed in different astrophysical systems and the resulting variation in temporal units, the unit time is specified in terms of a variable magnetic field.}}
    \label{fig:1e-6_escape}
\end{figure}

In Fig.~\ref{fig:1e-6_escape}, we show the evolution of the distribution function including particle escape and acceleration along with synchrotron loss. The escape time is controlled by the parameter $b = 10^{-6}$ which is kept fixed for the curves shown with different $a$ values in three different panels. 
For all the plots the high-energy cut-off show a rapid decrement towards lower $\gamma$ as compared to Fig.~\ref{fig:different_a} owing to the particle escape in addition to the synchrotron loss process.
One interesting observation is that due to escape, the evolution of the distribution begins with the movement of the high-energy cut-off to lower $\gamma$ and increasing the rate of the evolution towards the steady state.
After attaining the steady state the movement of the cut-off towards lower $\gamma$ ceases and the height of the distribution starts to decrease as a result of the particle escape. 
Such kind of evolution is a manifestation of the escape time-scale which follows $\gamma^{-{8}/{3}}$ implying that the high-energy particles have lower escape time and therefore they leaks out of the system faster than the low energy particles. 
We further show the evolution of the distribution function for $b=10^{-4}$ and $10^{-5}$ with different $a$ values in Figs.~\ref{fig:1e-5_escape}, \ref{fig:1e-4_escape} which exhibit an almost similar evolutionary dynamics.
However, due to lesser escape timescale than with $b=10^{-6}$ the particles leaks out much faster for the cases presented in Figs.~\ref{fig:1e-5_escape}, \ref{fig:1e-4_escape}.

{In summary, we investigate the impact of small-scale magnetohydrodynamic (MHD) fluctuations on the acceleration of high-energy particles in astrophysical environments. 
We find that the acceleration of high-energy protons is significantly enhanced by small-scale MHD fluctuations compared to electrons.
Turbulent acceleration mediated by these fluctuations enables protons to maintain their energy levels in the presence of radiative synchrotron cooling. 
Additionally, the inclusion of the spatial escape term, which arises from the parallel diffusion coefficient along the mean magnetic field, demonstrates a relatively dominant escape of higher energy particles from the system.}

%%%%%%%%%%%%%%%%%%%%%%%%%%%%%%%%%%%%%%%%%%%%%%
%Astrophysical Application
%%%%%%%%%%%%%%%%%%%%%%%%%%%%%%%%%%%%%%%%%%%%%%

\section{Astrophysical applications} \label{sec:application}
\subsection{Particle transport in the vicinity of relativistic shocks}
Small-scale turbulence has been identified as the primary scattering agent of non-thermal cosmic ray (CR) particles in the vicinity of relativistic shock waves, which is crucial for efficient Fermi acceleration \citep{lemoine_2006b}. 
The downstream medium experiences intense small-scale turbulence due to weakly magnetized upstream regions, providing efficient scattering of CR particles and aiding in the completion of enough Fermi-cycles to facilitate acceleration \citep{plotnikov_2011}. 
The resulting scenario leads to a power-law-like particle spectrum at the shock front, while small-scale turbulence in the downstream region could act as an agent to energize CR particles via stochastic turbulent acceleration as they move further downstream.

Resonant scattering of particles with turbulent waves in the downstream region, which is typical for low rigidity particles ($R_{l}<l_{c}$), has been examined as a second-order turbulent acceleration mechanism in parallel relativistic shock \citep{virtanen_2005} considering the quasilinear condition. 
It has been concluded that this mechanism could have a significant impact on the evolution of the particle spectrum. 
However, as observed by \cite{chang_2008,plotnikov_2011}, for a certain period of time, the downstream turbulence is expected to be mediated by intense ($\delta B\gg B_{0}$) small-scale magnetic turbulence, which decays with a damping rate $\propto k^{3}$ suggesting a comparative quick decay of the power available at small-scales.
Whereas larger modes show a damping rate of $\propto k^{2}$ indicating that the power on scales exceeding the Larmor radius of the bulk plasma decays on long, MHD scales \citep{keshet_2009, sironi_2015}.
 This observation implies that turbulent acceleration at the low rigidity regime with longer magnetohydrodynamic (MHD) modes could be dominant at later stages, but initially, charged particles will experience acceleration through small-scale turbulence.
 This argument is in consonance with the theory that small-scale turbulence leads to large-scale turbulence through an inverse cascade \citep{medvedev_2004,Katz_2007}, although further research is needed in this area.

Moreover, evidence of microturbulence generated through Weibel instability has been observed in the precursor region of relativistic shocks, where the upstream medium shows elongated filamentary structures \citep{plotnikov_2013}. 
This region provides a scenario where turbulent acceleration of charged particles could take place through small-scale turbulence.

This study focuses on the effect of stochastic turbulent acceleration on non-thermal CRs in the presence of small-scale turbulence, where no power is available at the scale of the gyro-radius of the particle.
However, the present study relies on quasilinear theory, and the intense small-scale microturbulence needed in relativistic shocks both upstream and downstream requires a larger turbulence intensity $\delta B\gg B_{0}$, which is not possible to achieve through the present analytical framework.
This study can be considered as an initial step to explore the turbulent acceleration mechanism in such intense turbulence scenarios. 
Although the study cannot provide a realistic quantification of the turbulent acceleration taking place in such microturbulence, the universality in the momentum diffusion coefficient gives a hint of the enriched physics, which would be interesting to explore and would be taken up in future works.

\subsection{Ballistic transport of cosmic ray in Blazars}
The regime under investigation in this work is commonly also referred to as the "Ballistic Regime" \citep{reichherzer_2022}, in which the parallel transport of charged particles is minimally affected. 
The Ballistic Regime is especially suitable for modeling the transportation of particles with extremely high energies, near the Hillas limit. 
It is also effective during the initial stages of particle acceleration, before the particle has spent enough time in the acceleration region for its transport to become diffusive \citep{reichherzer_2022_proceeding}.
Recent studies have attributed the transport of particles in the Ballistic Regime to explaining the spectral energy distributions and light curves of high-energy emission from Blazars \citep{tjus_2022a,tjus_2022b,reichherzer_2022_proceeding}. 
These works mainly focused on the spatial transport of high energetic particles in AGN-plasmoids, which are often speculated to be responsible for the observed temporal variability in the Blazar sources.

Recent studies have shown that cosmic ray particles with energies above a certain threshold are expected to follow Ballistic transport. 
Specifically, it has been suggested by \cite{tjus_2022a} that cosmic ray particles with energies above $E\gtrsim 5l_{c}cqB_{0}/2\pi$ are expected to exhibit Ballistic transport. 
For AGN-plasmoids, it has been found that the transport of protons with energies $\gtrsim 10^{15}$,eV should also be considered under the Ballistic regime. 
Such protons typically possess a Lorentz factor of the order of $10^{6}$.

In Fig.~\ref{fig:no_escape_proton}, it has been shown that small-scale turbulence can effectively provide continuous acceleration to such high-energy protons, enabling them to maintain their energy levels despite catastrophic radiative cooling. 
However, it is important to note that a more precise quantitative analysis is necessary to fully comprehend the transport of such particles in the context of AGN-plasmoids.
We believe that our work will be relevant in studying the effect of turbulent acceleration on cosmic ray particles in Ballistic transport regimes. 
In particular, it would be interesting to investigate the interplay between turbulent acceleration due to small-scale fluctuations and synchrotron loss in the context of AGN-plasmoids and Blazar variability.

%%%%%%%%%%%%%%%%%%%%%%%%%
%Discussion
%%%%%%%%%%%%%%%%%%%%%%%%%
\section{Summary and outlook}\label{sec:diss}

In this work we consider the effect of small-scale turbulence with a guided magnetic field on the acceleration of very high-energy charged particles.
This study is motivated primarily by an academic interest where our emphasis is towards understanding the nature of interaction between the cosmic rays and the stochastic magnetic fields at scales smaller than the particles' gyro-radius corresponding to the uniform guided magnetic field.
It is likely that such kind of situations occur in various astrophysical scenarios, for example in the vicinity of relativistic shocks.
We carry out a semi-analytic study based on the quasilinear theory of plasma and determine the momentum transport coefficients for scenarios involving both isotropic and anisotropic turbulence at small scales.
For Alfv\`{e}nic turbulence, we consider isotropic single-scale turbulence injection spectrum and anisotropic turbulence with cascade along $k_{\perp}$ direction. 
Our calculation indicates that in both the turbulence scenarios the transport coefficients follow an inverse power-law trend with the energy, or in other words, the Lorentz factor $\gamma$, of the charged particles.
In the present work, we obtain the following power law scaling relations for the turbulent transport coefficients: $D_{\gamma\gamma}\propto\gamma^{-2/3}$, $D_{\mu\mu}\propto\gamma^{-8/3}$, and $D_{\mu\gamma}=0$.
The earlier work by \cite{tsytovich_1977} reports a power law behaviour of $D_{\gamma \gamma}\propto \gamma^{-1}$ which is different from what we observe here.
Additionally, a similar trend for the transport coefficient $D_{\gamma\gamma}$ is found for fast magnetosonic wave turbulence with an isotropic single-scale injection spectrum, as described in detail in Appendix~\ref{sec:fast}. 
The similarity in the behavior of $D_{\gamma\gamma}$ for isotropic Alfv\`{e}n, anisotropic Alfv\'{e}n, and isotropic fast wave turbulence suggests that the behavior of the momentum diffusion coefficient becomes universal when turbulent fluctuations occur in the sub-gyro scale regime. 
Such universality of the diffusion coefficient has also been reported for spatial diffusion of charged particles in small-scale turbulent environment\footnote[1]{See \cite{plotnikov_2011} for small-scale magnetic field following white noise and parallel diffusion scaling as rigidity squared. 
Also, see \cite{subedi_2017} and \cite{dundovic_2020} for similar scaling in isotropic spatial diffusion in synthetically constructed turbulence fields. 
\cite{Pezzi_2022} showed the same trend for MHD turbulence.}.
Moreover, \cite{Pezzi_2022} also observed that the impact of anisotropy in MHD turbulence spectrum on the particle diffusion coefficient is weak, which we also observe for the momentum diffusion coefficient.
Therefore it appears that for particles with larger gyro-radius or greater energy, the specific details of the small-scale turbulence may not significantly affect particle distribution, though the presence of power at these scales can still impact the overall distribution.
The calculated form for $D_{\mu \mu}$ leads to a parallel spatial diffusion coefficient which scales with $\gamma$ as $\sim\gamma^{8/3}$; see Eq.~(\ref{eq:kas}) for $\mathcal{K}$.
Such a trend is compatible with the QLT constraint defined in Eq.~(\ref{eq:QLT}).
Moreover, the parallel diffusion that we find is similar to that of obtained from numerical simulation by \cite{cesse_2001}.
Their result showed that the parallel diffusion in quasilinear regime scales with rigidity (or $\gamma$) with an exponent of $7/3$ which is close to the result that we obtain from asymptotic expansion of the quasilinear diffusion coefficients.

Having {observed the trend of the} transport coefficients, we then solve the transport equation for the cosmic rays, i.e., the Fokker-Planck equation.
When we ignore the synchrotron loss and diffusive escape mechanisms, we find that the small-scale turbulence leads to the energization of particles in such a way that the low energy particles are accelarated faster compared to the particles with higher $\gamma$, {which is typical for Fermi type stochastic acceleration}. 
{Thus} resulting in a non-uniform acceleration of the cosmic ray particles.
A qualitative comparison with the case where the gyro-radius of particle is smaller than the turbulence correlation length reveals that the acceleration due to small-scale turbulence is relatively less efficient.

We further demonstrate the interplay of various micro-physical processes, such as acceleration, synchrotron loss and diffusive particle escape, on the particle spectrum in various regimes of the parameters $a$ and $b$ in Eq.~(\ref{eq:transport}).
The particle spectrum is observed to attain a steady-state as a result of the competition between acceleration and synchrotron loss. 
Additionally, with the particle escape process included, the distribution function
evolves in such a way that the high-energy particles leak out of the system faster than the low-energy ones.

Our study indicates that in a situation where small-scale turbulence is mediated by a mean magnetic field, acceleration of high energy particles would be small in presence of other more dominant competing micro-physical processes, such as the synchrotron loss and diffusive escape.
However, such an acceleration is found to be significant for lower energy particles.
The hump like structure that develops in the distribution function signifies the acceleration of low-$\gamma$ particles.
{Additionally, while investigating such interplay with electron and proton distributions individually, we observe that small-scale turbulence can accelerate protons to higher energies than electrons, and this acceleration may assist high-energy proton particles in maintaining their energy from synchrotron loss effects.}
{Therefore, w}e envisage that in some cases with appropriate values for the dynamical quantities this acceleration could become significant and it may help the non-thermal particles to sustain their energy against the radiative cooling mechanisms.
{Finally we discuss about adequate astrophysical systems that could provide suitable conditions for small-scale turbulence to potentially influence the energy distribution of non-thermal particles.}

This work is {the} first step towards studying a more complex interplay of various micro-physical processes and their impact on the energy spectrum of the cosmic ray.
Other acceleration scenarios where resonant interaction between turbulent waves and high energy cosmic ray particles is not included, for example, adiabatic acceleration due to random velocity of the MHD fluid \citep{ptuskin_1988,lemoine_2019} could also provide significant acceleration to the non-thermal particles with high rigidity.
Such acceleration configurations along with various energy loss processes will be considered in future works. 
As an extension to this present work, it would be interesting to see the effects of small-scale turbulence on the high energy non-thermal particles in a non-linear framework of wave-particle interaction \citep{beresnyak_2011,yan_2008}.
\section{Acknowledgement}
We would like to thank the anonymous referees for the helpful comments, and constructive remarks on this manuscript. 
S.K. and B.V. would like to thank the financial support from the Max Planck partner group award at Indian Institute of Technology, Indore.
S.K. further acknowledges support from STFC through grant ST/X001067/1.
Additionally, S.K. extends appreciation to Suchismita Banerjee for her assistance in typesetting the mathematical formulas presented here.

\section*{DATA AVAILABILITY}
The codes used in this work are available from the corresponding author upon reasonable request.

%%%%%%%%%%%%%%%%%%%%%%%% 
%REFERENCES 
%%%%%%%%%%%%%%%%%%%%%%%%

\bibliographystyle{mnras}
\bibliography{sst} 
\appendix

%%%%%%%%%%%%%%%%%%%%%%%%%
%Appendix
%%%%%%%%%%%%%%%%%%%%%%%%%
\section{Calculation of various turbulent correlation terms for MHD turbulence}
\label{sec:appendix1}
Here we show the derivations pertaining to the calculations of $T_{ij}$, $Q_{ij}$ and $R_{ij}$ for MHD turbulence. 
The Ohm's Law for MHD regime can be written in the following way,
\begin{equation*}
    \vec{E}\left(\vec{k}\right) = -\left(\frac{1}{c}\right)\vec{u}\left(\vec{k}\right)\times \vec{B}_{0}
    \implies E_{i}\left(\vec{k}\right) = -\left(\frac{1}{c}\right)\epsilon_{imn}u_{m}\left(\vec{k}\right)\times B_{0n}
\end{equation*}
Therefore, the electric field correlations can be written as follows,
\begin{equation*}
 \left\langle E_{i}\left(\vec{k}\right)E_{j}^{*}\left(\vec{k}'\right) \right\rangle = \frac{1}{c^2} \left\langle \epsilon_{imn}\epsilon_{jpq}u_{m}\left(\vec{k}\right)B_{0n}u_{p}^{*}\left(\vec{k}'\right)B_{0q} \right\rangle
\end{equation*}
As mean B-field is only directed along z direction, the correlation becomes,
\begin{equation*}
   \left\langle E_{i}\left(\vec{k}\right)E_{j}^{*}\left(\vec{k}'\right) \right\rangle = \frac{B_{0}^2}{c^2} \left\langle \epsilon_{imz}\epsilon_{jpz}u_{m}\left(\vec{k}\right)u_{p}^{*}\left(\vec{k}'\right) \right\rangle  
\end{equation*}
Employing the identity $\epsilon_{imn}\epsilon_{jpq}=\left(\delta_{ij}\delta_{mp}-\delta_{ip}\delta_{jm}\right)$ the correlation function simplifies to, 
% \begin{equation*}
%   \implies \left\langle E_{i}\left(\vec{k}\right)E_{j}^{*}\left(\vec{k}'\right) \right\rangle = \frac{B_{0}^2}{c^2} \left\langle \left(\delta_{ij}\delta_{mp}-\delta_{ip}\delta_{jm}\right)v_{m}\left(\vec{k}\right)v_{p}^{*}\left(\vec{k}'\right) \right\rangle  
% \end{equation*}
\begin{equation}\label{eq:mhd_corr}
\left\langle E_{i}\left(\vec{k}\right)E_{j}^{*}\left(\vec{k}'\right) \right\rangle = \frac{B_{0}^2}{c^2} \left[\delta_{ij}\left\langle u_{p}\left(\vec{k}\right)u_{p}^{*}\left(\vec{k}'\right) \right\rangle - \left\langle u_{j}\left(\vec{k}\right)u_{i}^{*}\left(\vec{k}'\right) \right\rangle \right]
\end{equation}
Similarly, the electric field magnetic field correlation becomes,
\begin{equation}
  \left\langle E_{i}\left(\vec{k}\right)B_{j}^{*}\left(\vec{k}'\right) \right\rangle = -\frac{1}{c} \left\langle \epsilon_{imn}u_{m}\left(\vec{k}\right)B_{0n}B_{j}^{*}\left(\vec{k}'\right) \right\rangle = - \frac{B_{0z}}{c} \left\langle \epsilon_{imz}u_{m}\left(\vec{k}\right)B_{j}^{*}\left(\vec{k}'\right)\right\rangle
\end{equation}
and,
\begin{equation*}
  \left\langle B_{i}\left(\vec{k}\right)E_{j}^{*}\left(\vec{k}'\right) \right\rangle = - \frac{B_{0z}}{c} \left\langle B_{i}\left(\vec{k}\right)\epsilon_{ipz}u_{p}^{*}\left(\vec{k}'\right)\right\rangle = - \frac{B_{0z}}{c} \epsilon_{ipz} \left\langle B_{i}\left(\vec{k}\right)u_{p}^{*}\left(\vec{k}'\right)\right\rangle
\end{equation*}
Following the definitions given in Eq.~(\ref{eq:red_corr}) we get,
\begin{enumerate}
    \item $ \DS\left\langle E_{i}\left(\vec{k}\right)E_{j}^{*}\left(\vec{k}'\right) \right\rangle =  \frac{B_{0}^2}{c^2} \left[\delta_{ij}V_{A}^2\delta\left(\vec{k}-\vec{k}'\right)Y_{pp}\left(k\right)-V_{A}^2\delta\left(\vec{k}-\vec{k}'\right)Y_{ji}\left(\vec{k}\right)\right] = \frac{B_{0}^{2}V_{A}^{2}}{c^2}\left[\delta_{ij}Y_{pp}\left(\vec{k}\right)-Y_{ji}\left(\vec{k}\right)\right]$
     \item $ \DS\left\langle E_{i}\left(\vec{k}\right)B_{j}^{*}\left(\vec{k}'\right) \right\rangle = -\frac{B_{0}}{c} \epsilon_{imz}\left\langle u_{m}\left(\vec{k}\right)B_{j}^{*}\left(\vec{k}'\right) \right\rangle = -\frac{B_{0}^{2}V_{A}}{c} \epsilon_{imz}\delta\left(\vec{k}-\vec{k}'\right)C_{mj}\left(\vec{k}\right)$
      \item $ \DS\left\langle B_{i}\left(\vec{k}\right)E_{j}^{*}\left(\vec{k}'\right) \right\rangle = - \frac{B_{0}}{c} \epsilon_{jpz} \left\langle B_{i}\left(\vec{k}\right)u_{p}^{*}\left(\vec{k}'\right)\right\rangle = -\frac{B_{0}^{2}V_{A}}{c}\epsilon_{jpz}\delta\left(\vec{k}-\vec{k}'\right)C_{ip}\left(\vec{k}\right)$
\end{enumerate}

\section{Calculation of correlation functions}
\label{sec:appendix2}
In this appendix, we derive the transformation laws for various turbulent spectra, from Cartesian space to polarization space.
\begin{eqnarray}
\nonumber
    R_{11}\delta(k-k') &=& \left\langle E_{1}(k) E_{1}^{*}(k') \right\rangle 
    = \left\langle \frac{E_{\mathcal{R}}(k)+E_{\mathcal{L}}(k)}{\sqrt{2}}.\frac{E_{\mathcal{R}}^{*}(k')+E_{\mathcal{L}}^{*}(k')}{\sqrt{2}} \right\rangle \\
    &=& \frac{1}{2}\left[ \left\langle E_{\mathcal{R}}(k)E_{\mathcal{R}}^{*}(k')\right\rangle + \left\langle E_{\mathcal{R}}(k)E_{\mathcal{L}}^{*}(k')\right\rangle + \left\langle E_{\mathcal{L}}(k)E_{\mathcal{R}}^{*}(k')\right\rangle + \left\langle E_{\mathcal{L}}(k)E_{\mathcal{L}}^{*}(k')\right\rangle\right]
\end{eqnarray}
Where, $E_{\mathcal{R}}$ and $E_{\mathcal{L}}$ are known as Jones vectors and they are defined as, 
$$E_{\mathcal{R}}=\frac{E_{1}-\iota E_{2}}{\sqrt{2}},\quad\quad E_{\mathcal{L}}=\frac{E_{1}+\iota E_{2}}{\sqrt{2}}$$. 
$$\implies E_{1} = \frac{E_{\mathcal{R}}+E_{\mathcal{L}}}{\sqrt{2}},\quad\quad E_{2} = \frac{E_{\mathcal{L}}-E_{\mathcal{R}}}{\iota\sqrt{2}} = \frac{\iota(E_{\mathcal{R}}-E_{\mathcal{L}})}{\sqrt{2}}$$.
Further with the following definitions of the correlation tensor in the polarization space $R_{11}$ becomes,
\begin{itemize}
    \item $\DS R_{\mathcal{RR}}\delta(k-k') = \left\langle E_{\mathcal{R}}(k) E_{\mathcal{R}}^{*}(k') \right\rangle$;
    \item $\DS R_{\mathcal{RL}}\delta(k-k') = \left\langle E_{\mathcal{R}}(k) E_{\mathcal{L}}^{*}(k') \right\rangle$;
    \item $\DS R_{\mathcal{LR}}\delta(k-k') = \left\langle E_{\mathcal{L}}(k) E_{\mathcal{R}}^{*}(k') \right\rangle$;
    \item  $\DS R_{\mathcal{LL}}\delta(k-k') = \left\langle E_{\mathcal{L}}(k) E_{\mathcal{L}}^{*}(k') \right\rangle$;
\end{itemize}
$$R_{11} = \frac{1}{2}\left(R_{\mathcal{RR}}+R_{\mathcal{RL}}+R_{\mathcal{LR}}+R_{\mathcal{LL}}\right)$$
Similarly $R_{22}$, $R_{12}$ and $R_{21}$ can be written as, 
\begin{eqnarray}
\nonumber
    R_{22}\delta(k-k') &=& \left\langle E_{2}(k) E_{2}^{*}(k') \right\rangle 
    = \frac{1}{2}\left[ \left\langle E_{\mathcal{R}}(k)E_{\mathcal{R}}^{*}(k')\right\rangle + \left\langle E_{\mathcal{L}}(k)E_{\mathcal{L}}^{*}(k')\right\rangle - \left\langle E_{\mathcal{R}}(k)E_{\mathcal{L}}^{*}(k')\right\rangle - \left\langle E_{\mathcal{L}}(k)E_{\mathcal{R}}^{*}(k')\right\rangle\right] \\
    &=& -\frac{1}{2}\left(R_{\mathcal{RR}}+R_{\mathcal{LL}}-R_{\mathcal{LR}}-R_{\mathcal{RL}}\right)\delta(k-k').
\end{eqnarray}
\begin{eqnarray}
\nonumber
    R_{12}\delta(k-k') &=& \left\langle E_{1}(k) E_{2}^{*}(k') \right\rangle 
    = \left\langle \frac{E_{\mathcal{R}}(k)+E_{\mathcal{L}}(k)}{\sqrt{2}} -\iota\frac{(E_{\mathcal{R}}^{*}(k')-E_{\mathcal{L}}^{*}(k'))}{\sqrt{2}}\right\rangle \\ \nonumber
    &=& -\frac{\iota}{2} \left[ \left\langle E_{\mathcal{R}}(k)E_{\mathcal{R}}^{*}(k')\right\rangle - \left\langle E_{\mathcal{R}}(k)E_{\mathcal{L}}^{*}(k')\right\rangle + \left\langle E_{\mathcal{L}}(k)E_{\mathcal{R}}^{*}(k')\right\rangle - \left\langle E_{\mathcal{L}}(k)E_{\mathcal{L}}^{*}(k')\right\rangle\right] \\
    &=& -\frac{\iota}{2}\left(R_{\mathcal{RR}}-R_{\mathcal{RL}}+R_{\mathcal{LR}}-R_{\mathcal{LL}}\right)\delta(k-k').
\end{eqnarray}
\begin{eqnarray}
    R_{21}\delta(k-k') &=& \left\langle E_{2}(k) E_{1}^{*}(k') \right\rangle =\frac{\iota}{2} \left[ \left\langle E_{\mathcal{R}}(k)E_{\mathcal{R}}^{*}(k')\right\rangle + \left\langle E_{\mathcal{R}}(k)E_{\mathcal{L}}^{*}(k')\right\rangle - \left\langle E_{\mathcal{L}}(k)E_{\mathcal{R}}^{*}(k')\right\rangle - \left\langle E_{\mathcal{L}}(k)E_{\mathcal{L}}^{*}(k')\right\rangle\right] \\
    &=& \frac{\iota}{2}\left(R_{\mathcal{RR}}+R_{\mathcal{RL}}-R_{\mathcal{LR}}-R_{\mathcal{LL}}\right)\delta(k-k').
\end{eqnarray}
Therefore, we can write,
\begin{eqnarray}
\begin{pmatrix}
R_{11}\\
R_{12}\\
R_{21}\\
R_{22}
\end{pmatrix} = \frac{1}{2} \begin{pmatrix}
1 & 1 & 1 & 1\\
-\iota & \iota & -\iota & \iota\\
\iota & \iota & -\iota & -\iota\\
1 & -1 & -1 & 1
\end{pmatrix} \begin{pmatrix}
R_{\mathcal{RR}}\\
R_{\mathcal{RL}}\\
R_{\mathcal{LR}}\\
R_{\mathcal{LL}}
\end{pmatrix}
\end{eqnarray}

The matrix upon inversion gives, 
\begin{itemize}
    \item $\DS R_{\mathcal{RR}} = \frac{1}{2}\left(R_{11}+R_{22}+\iota(R_{12}-R_{21})\right)$;
    \item $\DS R_{\mathcal{LL}} = \frac{1}{2}\left(R_{11}+R_{22}-\iota(R_{12}-R_{21})\right)$;
    \item $\DS R_{\mathcal{LR}} = \frac{1}{2}\left(R_{11}-R_{22}+\iota(R_{12}+R_{21})\right)$;
    \item $\DS R_{\mathcal{RL}} = \frac{1}{2}\left(R_{11}-R_{22}-\iota(R_{12}+R_{21})\right)$;
\end{itemize}
By definition (see Eq.~\ref{eq:corr}),
\begin{eqnarray}
    R_{ij}(k)\delta(k-k') &=& \left\langle E_{i}(k) E_{j}^{*}(k') \right\rangle
    = \frac{B_{0}^{2}}{c^{2}}\left(\delta_{ij} \left\langle u_{p}(k)u_{p}^{*}(k') \right\rangle - \left\langle u_{j}(k)u_{i}^{*}(k') \right\rangle \right)
    \\
    \implies R_{ij}
    &=& \frac{B_{0}^{2} V_{A}^{2}}{c^{2}}\left(\delta_{ij} Y_{pp}(k) - Y_{ji}(k)\right) 
    =\frac{B_{0}^{2} V_{A}^{2}}{c^{2}}\left(\delta_{ij} Y_{pp} - Y_{ij} \right)
    = \frac{B_{0}^{2} V_{A}^{2}}{c^{2}k^{2}}\left(\delta_{ij} + \frac{k_{i}k_{j}}{k^{2}}\right) P_{0}\delta\left(\frac{k}{m'k_{g}}-1\right)
\end{eqnarray}
where we have used Eq.~(\ref{eq:mhd_corr}) for the electric field correlation term, Eq.~(\ref{eq:turb_spec}) for the small-scale turbulence spectrum and utilized its symmetry property ($Y_{ij} = Y_{ji}$).
We further use the following identity for $Y_{pp}$,
$$Y_{pp}\left(\vec{k}\right) = k^{-2}\left(\delta_{pp}-1\right)P_{0}\delta\left(\frac{k}{m'k_{g}}-1\right)= \frac{2P_{0}}{k^{2}}\delta\left(\frac{k}{m'k_{g}}-1\right).$$
Therefore, with the definition of $\vec{k}= \{k_{\perp}\cos{\psi},k_{\perp}\sin{\psi},k_{||}\}$ the correlations in the polarization space takes the following form, 
\begin{itemize}
    \item $\DS R_{\mathcal{RR}} = \frac{1}{2} \frac{B_{0}^{2} V_{A}^{2}}{c^{2}k^{2}}P_{0}\delta\left(\frac{k}{m'k_{g}}-1\right)\left[1+\frac{k_{\perp}^{2}\cos{\psi}^{2}}{k^{2}}+1+\frac{k_{\perp}^{2}\sin{\psi}^{2}}{k^{2}}+\iota\left(\frac{k_{\perp}^{2}}{k^{2}}\cos{\psi}\sin{\psi}-\frac{k_{\perp}^{2}}{k^{2}}\cos{\psi}\sin{\psi}\right)\right]$
    
    \hskip7.5ex$\DS = \frac{B_{0}^{2} V_{A}^{2}}{c^{2}k^{2}}P_{0}\left(1+\frac{k_{\perp}^{2}}{2k^{2}}\right)\delta\left(\frac{k}{m'k_{g}}-1\right)$;
    \item $\DS R_{\mathcal{LL}} = \frac{B_{0}^{2} V_{A}^{2}}{c^{2}k^{2}}P_{0}\left(1+\frac{k_{\perp}^{2}}{2k^{2}}\right)\delta\left(\frac{k}{m'k_{g}}-1\right)$;
    \item $\DS R_{\mathcal{LR}} = \frac{1}{2} \left( 1+\frac{k_{\perp}^{2}\cos{\psi}^{2}}{k^{2}}-1-\frac{k_{\perp}^{2}\sin{\psi}^{2}}{k^{2}}+\iota\left(2\frac{k_{\perp}^{2}}{k^{2}}\cos{\psi}\sin{\psi}\right)\right) \frac{B_{0}^{2} V_{A}^{2}}{c^{2}k^{2}}P_{0}\delta\left(\frac{k}{m'k_{g}}-1\right)= \frac{1}{2} \frac{B_{0}^{2} V_{A}^{2}}{c^{2}k^{2}}P_{0}\delta\left(\frac{k}{m'k_{g}}-1\right)\frac{k_{\perp}^{2}}{k^{2}} e^{2\iota\psi}$
    \item $\DS R_{\mathcal{RL}}= \frac{1}{2} \frac{B_{0}^{2} V_{A}^{2}}{c^{2}k^{2}}P_{0}\delta\left(\frac{k}{m'k_{g}}-1\right)\frac{k_{\perp}^{2}}{k^{2}} e^{-2\iota\psi}$
\end{itemize}
where we have used the following,
\begin{eqnarray}
    \delta_{ij}+\frac{k_{i}k_{j}}{k^2} = 
    \begin{bmatrix}
    1 & 0 & 0 \\
    0 & 1 & 0 \\
    0 & 0 & 1
    \end{bmatrix} + \frac{1}{k^2}
    \begin{bmatrix}
    k_{\perp}^2\cos^2{\psi} & k_{\perp}^2\cos{\psi}\sin{\psi} & k_{\perp}k_{||}\cos{\psi} \\
    k_{\perp}^2\cos{\psi}\sin{\psi} & k_{\perp}^2\sin^2{\psi} & k_{\perp}k_{||}\sin{\psi} \\
    k_{||}k_{\perp}\cos{\psi} & k_{||}k_{\perp}\sin{\psi} & k_{||}^2
    \end{bmatrix}
\end{eqnarray}
Similarly considering $P_{ij}=B_{0}^2 Y_{ij}$ we get, 
\begin{itemize}
    \item $\DS P_{\mathcal{RR}}=B_{0}^2 Y_{\mathcal{RR}} = \frac{B_{0}^2}{k^{2}} \left(1-\frac{k_{\perp}^2}{2k^2}\right)P_{0}\delta\left(\frac{k}{m'k_{g}}-1\right)$;
    \item $\DS P_{\mathcal{LL}}= B_{0}^2 Y_{\mathcal{LL}} = \frac{B_{0}^2}{k^{2}} \left(1-\frac{k_{\perp}^2}{2k^2}\right)P_{0}\delta\left(\frac{k}{m'k_{g}}-1\right)$;
    \item $\DS P_{\mathcal{LR}}=B_{0}^2 Y_{\mathcal{LR}} = \frac{-B_{0}^2}{2}\frac{k_{\perp}^2}{k^4}e^{2\iota\psi}P_{0}\delta\left(\frac{k}{m'k_{g}}-1\right)$;
    \item $\DS P_{\mathcal{RL}}=B_{0}^2 Y_{\mathcal{RL}} = \frac{-B_{0}^2}{2}\frac{k_{\perp}^2}{k^4}e^{-2\iota\psi}P_{0}\delta\left(\frac{k}{m'k_{g}}-1\right)$.
\end{itemize}
Further, following the definition of $T_{ij}$ and $C_{ij}$ while noting the fact that $C_{ij} = \sigma Y_{ij}$ and $Y_{ij}$ is real, the velocity-magnetic field correlation function becomes,
\begin{itemize}
    \item $\DS T_{11} = -\frac{B_{0}^2V_{A}}{c}\epsilon_{1m3}C_{m1}^{*}= - \frac{B_{0}^2V_{A}}{c}\epsilon_{123}C_{21}^{*}= - \frac{B_{0}^{2}V_{A}}{c}C_{21}^{*}=\frac{B_{0}^2 V_{A}}{c}\sigma \frac{k_{\perp}^2}{k^4}\cos{\psi}\sin{\psi}P_{0}\delta\left(\frac{k}{m'k_{g}}-1\right)$;
    \item $\DS T_{22} = - \frac{B_{0}^2V_{A}}{c}\epsilon_{2m3}C_{m2}^{*}= \frac{B_{0}^{2}V_{A}}{c}C_{12}^{*}=-\frac{B_{0}^2 V_{A}}{c}\sigma \frac{k_{\perp}^2}{k^4}\cos{\psi}\sin{\psi}P_{0}\delta\left(\frac{k}{m'k_{g}}-1\right)$;
    \item $\DS T_{12} = - \frac{B_{0}^2V_{A}}{c}\epsilon_{2m3}C_{m1}^{*}= - \frac{B_{0}^{2}V_{A}}{c}\epsilon_{213}C_{11}^{*}= \frac{B_{0}^{2}V_{A}}{c}C_{11}^{*}=\frac{B_{0}^2 V_{A}}{ck^{2}}\sigma \left(1-\frac{k_{\perp}^2\cos{\psi}^2}{k^2}\right)P_{0}\delta\left(\frac{k}{m'k_{g}}-1\right)$
    \item $\DS T_{21} = - \frac{B_{0}^2V_{A}}{c}\epsilon_{1m3}C_{m2}^{*}=- \frac{B_{0}^{2}V_{A}}{c}C_{22}^{*}=-\frac{B_{0}^2 V_{A}}{ck^{2}}\sigma \left(1-\frac{k_{\perp}^2\sin{\psi}^2}{k^2}\right)P_{0}\delta\left(\frac{k}{m'k_{g}}-1\right)$
\end{itemize}
where $\epsilon_{ijk}$ is the Levi-Civita symbol and we have further used the fact that $C_{ij}=C^{*}_{ij}$.
Following these the correlation in the polarization space takes the form,
\begin{itemize}
    \item $\DS T_{\mathcal{RR}} = \frac{P_{0}}{2k^{2}}\left(\iota\frac{B_{0}^2 V_{A} \sigma}{c}\left(1-\frac{k_{\perp}^2\cos{\psi}^2}{k^2}+1-\frac{k_{\perp}^2\sin{\psi}^2}{k^2}\right) \right)\delta\left(\frac{k}{m'k_{g}}-1\right)
    = \iota\frac{B_{0}^2 V_{A} \sigma}{2ck^{2}}\left(2-\frac{k_{\perp}^2}{k^2}\right)P_{0}\delta\left(\frac{k}{m'k_{g}}-1\right)$;
    \item $\DS T_{\mathcal{LL}} = -\iota\frac{B_{0}^2 V_{A} \sigma}{2ck^{2}}\left(2-\frac{k_{\perp}^2}{k^2}\right)P_{0}\delta\left(\frac{k}{m'k_{g}}-1\right)$;
    \item $\DS T_{\mathcal{LR}} = \frac{P_{0}}{k^{2}} \left[\frac{B_{0}^2 V_{A} \sigma}{2c}\left(\frac{k_{\perp}^2\cos{\psi}\sin{\psi}}{k^2}+\frac{k_{\perp}^2\cos{\psi}\sin{\psi}}{k^2} \right)    +\iota\left(1-\frac{k_{\perp}^2\cos{\psi}^2}{k^2}-1+\frac{k_{\perp}^2\sin{\psi}^2}{k^2} \right)\right]\delta\left(\frac{k}{m'k_{g}}-1\right)$
    
    \hskip7ex $\DS =-\iota\frac{B_{0}^2 V_{A} \sigma}{2c}\frac{k_{\perp}^2}{k^4} e^{2\iota\psi}P_{0}\delta\left(\frac{k}{m'k_{g}}-1\right)$;
    \item $\DS T_{\mathcal{RL}}= \iota \frac{B_{0}^2 V_{A} \sigma}{2c}\frac{k_{\perp}^2}{k^4} e^{-2\iota\psi}P_{0}\delta\left(\frac{k}{m'k_{g}}-1\right)$.
\end{itemize}
Note that all the velocity-magnetic field correlations becomes imaginary in the polarization space.

\section{Derivation of $D_{pp}$}\label{sec:appendix3}
From Eq.~(\ref{eq:Dppinitial}) we get,
\begin{eqnarray}
\nonumber
    D_{pp} = \frac{\Omega^2\left(1-\mu^2\right)}{2}m^2c^2\frac{V_{A}^{2}}{c^{2}}P_{0}\mathcal{R}e\left[\sum_{n = -\infty}^{\infty}\int_{k_{min}}^{k_{max}}\delta\left(\frac{k}{m'k_{g}}-1\right)k^{-2}\,d^3\vec{k} \int_{0}^{\infty}dt     e^{-\iota\left(k_{||}v_{||}-\omega+n\Omega\right)t}
    \left\{\left(J_{n+1}^2\left(\frac{k_{\perp}v_{\perp}}{\Omega}\right)+J_{n-1}^2\left(\frac{k_{\perp}v_{\perp}}{\Omega}\right)\right)\left(1+\frac{k_{\perp}^{2}}{2k^{2}}\right) \right .\right.
    \\ \left .\left. 
    +J_{n+1}\left(\frac{k_{\perp}v_{\perp}}{\Omega}\right)J_{n-1}\left(\frac{k_{\perp}v_{\perp}}{\Omega}\right)\frac{k_{\perp}^{2}}{k^{2}} \right\}\right]
\end{eqnarray}
Time integration leads,
\begin{eqnarray}
\nonumber
    D_{pp} = \frac{\Omega^2\left(1-\mu^2\right)}{2}m^2c^2\frac{V_{A}^{2}}{c^{2}}\pi P_{0}\mathcal{R}e\left[\sum_{n = -\infty}^{\infty}\int_{k_{min}}^{k_{max}}\delta\left(\frac{k}{m'k_{g}}-1\right)\,dk \int_{0}^{\pi}\sin{\theta}d\theta \int_{0}^{2\pi}d\phi\,\delta(k\cos{\theta}v_{||}-\omega+n\Omega)\right. \\
    \left.
    \left\{\left(J_{n+1}^2\left(\frac{k_{\perp}v_{\perp}}{\Omega}\right)+J_{n-1}^2\left(\frac{k_{\perp}v_{\perp}}{\Omega}\right)\right)\left(1+\frac{k_{\perp}^{2}}{2k^{2}}\right)  
    +J_{n+1}\left(\frac{k_{\perp}v_{\perp}}{\Omega}\right)J_{n-1}\left(\frac{k_{\perp}v_{\perp}}{\Omega}\right)\frac{k_{\perp}^{2}}{k^{2}} \right\}\right]
\end{eqnarray}
The presence of the $\delta$ function inside the $\phi$ integration, gives rise to a resonance condition which the plasma waves and the charged particles have to satisfy in order for interaction to happen between them. 
In this work, we consider the resonance to be exact with no broadening.
Additional modifications regarding the resonance condition is also considered in literature which results from modifications of the quasilinear approach \citep[see][for example]{yan_2008}.
Upon performing the $\phi$ integration we get,
\begin{eqnarray}
\nonumber
    D_{pp} = \Omega^2\left(1-\mu^2\right)m^2c^2\frac{V_{A}^{2}}{c^{2}}\pi^{2} P_{0}\mathcal{R}e\left[\sum_{n = -\infty}^{\infty}\int_{k_{min}}^{k_{max}}\delta\left(\frac{k}{m'k_{g}}-1\right)\,dk \int_{0}^{\pi}\sin{\theta}d\theta\,\delta(k\cos{\theta}v_{||}-\omega+n\Omega)\right. \\
    \left.
    \left\{\left(J_{n+1}^2\left(\frac{k_{\perp}v_{\perp}}{\Omega}\right)+J_{n-1}^2\left(\frac{k_{\perp}v_{\perp}}{\Omega}\right)\right)\left(1+\frac{k_{\perp}^{2}}{2k^{2}}\right)  
    +J_{n+1}\left(\frac{k_{\perp}v_{\perp}}{\Omega}\right)J_{n-1}\left(\frac{k_{\perp}v_{\perp}}{\Omega}\right)\frac{k_{\perp}^{2}}{k^{2}} \right\}\right].
\end{eqnarray}
Further, we note that $k_{\perp}=k\sin\theta=k\sqrt{1-x^{2}}$ with $x$ being the cosine of the angle between $\vec{k}$ and the direction of the mean magnetic field, $x=\cos\theta$.
Owing to the small scale limit, $\frac{k_{\perp}v_{\perp}}{\Omega} >> 1$, the summation on 'n' becomes integration and integration over delta function with resonance condition leads \citep{tsytovich_1977},
\begin{eqnarray}
\nonumber
    D_{pp} \simeq \Omega\left(1-\mu^2\right)m^2c^2\frac{V_{A}^{2}}{c^{2}}\pi^{2} P_{0}\mathcal{R}e\left[\int_{0}^{\infty}\delta\left(\frac{k}{m'k_{g}}-1\right)\,dk \int_{-1}^{1}dx\,
    \left\{\left(J_{\frac{\omega}{\Omega}-\frac{kxv\mu}{\Omega}+1}^2\left(\frac{kv}{\Omega}\sqrt{1-x^2}\sqrt{1-\mu^2}\right)\right.\right.\right. \\
    \left.\left.\left.+J_{\frac{\omega}{\Omega}-\frac{kxv\mu}{\Omega}-1}^2\left(\frac{kv}{\Omega}\sqrt{1-x^2}\sqrt{1-\mu^2}\right)\right)\left(\frac{3-x^{2}}{2}\right)  
    +\left(1-x^{2}\right)J_{\frac{\omega}{\Omega}-\frac{kxv\mu}{\Omega}+1}\left(\frac{kv}{\Omega}\sqrt{1-x^2}\sqrt{1-\mu^2}\right)J_{\frac{\omega}{\Omega}-\frac{kxv\mu}{\Omega}-1}\left(\frac{kv}{\Omega}\sqrt{1-x^2}\sqrt{1-\mu^2}\right) \right\}\right],
\end{eqnarray}
where we consider $v_{\perp}=v\sqrt{1-\mu^{2}}$ with $\mu$ being the pitch-angle and the limit of the $k$ integration to be $0$ to $\infty$.
Performing the $k$ integration leads,
\begin{eqnarray}
\nonumber
    D_{pp} \simeq \Omega\left(1-\mu^2\right)m^2c^2\frac{V_{A}^{2}}{c^{2}}\pi^{2}m'k_{g} P_{0}\mathcal{R}e\left[ \int_{-1}^{1}dx
    \left\{\left(J_{\frac{\omega}{\Omega}-\frac{m'k_{g}xv\mu}{\Omega}+1}^2\left(\frac{m'k_{g}v}{\Omega}\sqrt{1-x^2}\sqrt{1-\mu^2}\right)\right.\right.\right. \\ \nonumber
    \left.\left.\left.
    +J_{\frac{\omega}{\Omega}-\frac{m'k_{g}xv\mu}{\Omega}-1}^2\left(\frac{m'k_{g}v}{\Omega}\sqrt{1-x^2}\sqrt{1-\mu^2}\right)\right)\left(\frac{3-x^{2}}{2}\right) \right.\right. \\ \left.\left.
    +\left(1-x^{2}\right)J_{\frac{\omega}{\Omega}-\frac{m'k_{g}xv\mu}{\Omega}+1}\left(\frac{m'k_{g}v}{\Omega}\sqrt{1-x^2}\sqrt{1-\mu^2}\right)J_{\frac{\omega}{\Omega}-\frac{m'k_{g}xv\mu}{\Omega}-1}\left(\frac{m'k_{g}v}{\Omega}\sqrt{1-x^2}\sqrt{1-\mu^2}\right) \right\}\right],
\end{eqnarray}

\section{Transport Equation} \label{sec:transport}
From Eq.~(\ref{eq:schlickeizer}) following $a_{1}=0$, due to the fact that $D_{\mu p}=0$, leads to
\begin{eqnarray}
    \frac{\partial F}{\partial t} =\frac{\partial}{\partial z}\left(\mathcal{K}\frac{\partial F}{\partial z}\right)-U\frac{\partial F}{\partial z} + \frac{p}{3}\frac{\partial U}{\partial z}\frac{\partial F}{\partial p} +\frac{1}{p^{2}}\frac{\partial}{\partial p}\left(p^{2}a_{2}\frac{\partial F}{\partial p}\right)+S_{0}
\end{eqnarray}
Substituting $F = \frac{f}{p^{2}}$, $f$ follows the following equation,
\begin{eqnarray}
    \frac{\partial f}{\partial t} =\frac{\partial}{\partial z}\left(\mathcal{K}\frac{\partial f}{\partial z}\right)-U\frac{\partial f}{\partial z} + \frac{p^{3}}{3}\frac{\partial U}{\partial z}\frac{\partial}{\partial p}\left(\frac{f}{p^2}\right) +\frac{\partial}{\partial p}\left(p^{2}a_{2}\frac{\partial}{\partial p}\left(\frac{f}{p^2}\right)\right)+S_{0}p^{2}
\end{eqnarray}
which upon simplification gets the following form,
\begin{eqnarray}
    \frac{\partial f}{\partial t} =\frac{\partial}{\partial z}\left(\mathcal{K}\frac{\partial f}{\partial z}\right)-\frac{\partial (Uf)}{\partial z}+ \frac{\partial}{\partial p}\left(\frac{\partial U}{\partial z}\frac{p}{3}f\right)+ \frac{\partial}{\partial p} \left(a_{2}\frac{\partial f}{\partial p}\right) - \frac{\partial}{\partial p}\left(\frac{2a_{2}f}{p}\right)+S_{0}p^{2}.
\end{eqnarray}
For the present work we neglect the $3^{rd}$ term on the right hand side which corresponds to adiabatic expansion loss and introduce a radiative loss term due to synchrotron process.
We further employ the leaky-box approximation \citep{lerche_1985}, following which we replace the spatial advection and diffusion terms by a momentum dependent escape term.
Such an approximation leads the transport equation to take the following form,
\begin{eqnarray}
    \frac{\partial f}{\partial t} + \frac{\partial}{\partial \gamma}\left(\frac{2Df}{\gamma}- c_{0}B^2\gamma^2 f\right) = \frac{\partial}{\partial \gamma}\left(D\frac{\partial f}{\partial \gamma}\right) - \frac{f}{T_{esc}}+ S_{0}\gamma^2.
\end{eqnarray}
Note that, the above equation is written in terms of particle's Lorentz factor $\gamma$ instead of momentum $p$ and the corresponding conversion factor is encapsulated within the constant factors of the transport coefficients.
Further, following the forms of the transport coefficients for small-scale turbulence as discussed in sections~\ref{sec:dpp} and \ref{sec:dmumu}, we find $1/T_{esc} = \alpha \gamma^{8/3}$, $D = D_{0} \gamma^{-2/3}$, with $\alpha$ and $\beta$ being the constants whose values depends on $\beta_{A}$, $m'$ and $\sigma$.
Upon substitution of the transport coefficients, the transport equation takes the following form,
\begin{eqnarray}
    \frac{\partial f}{\partial (T_{s}T)} + \frac{\partial}{\partial (\gamma_{s}\Gamma)}\left(\frac{2D_{0}(\gamma_{s}\Gamma)^{-\frac{2}{3}}f}{\gamma_{s}\Gamma}- c_{0}B^2(\gamma_{s}\Gamma)^2 f\right) =
    \frac{\partial}{\partial (\gamma_{s}\Gamma)}\left(D_{0} (\gamma_{s}\Gamma)^{-\frac{2}{3}}\frac{\partial f}{\partial (\gamma_{s}\Gamma)}\right) - f\alpha (\gamma_{s}\Gamma)^{\frac{8}{3}}+ S_{0}(\gamma_{s}\Gamma)^2.
\end{eqnarray}
Note that the substitution of the transport coefficients is done considering $t=T_{s}T$ and $\gamma=\gamma_{s}\Gamma$ with $T_{s}$ and $\gamma_{s}$ being the scaled time and Lorentz factor respectively. 
The above transport equation when written in the scaled units, simplifies to, 
\begin{eqnarray*}
    \frac{\partial f}{\partial T} + T_{s}\frac{\partial}{\partial \Gamma}\left\{\frac{2D_{0}(\gamma_{s}\Gamma)^{-\frac{5}{3}}f}{\gamma_{s}}- \frac{c_{0}B^2}{\gamma_{s}}(\gamma_{s}\Gamma)^2 f\right\} =
    T_{s}\frac{\partial}{\partial \Gamma}\left(\frac{D_{0}}{\gamma_{s}^2} (\gamma_{s}\Gamma)^{-\frac{2}{3}}\frac{\partial f}{\partial \Gamma}\right) - T_{s}f\alpha (\gamma_{s}\Gamma)^{\frac{8}{3}}+ T_{s}S_{0}(\gamma_{s}\Gamma)^2.
\end{eqnarray*}
Next, we consider $T_{s}$ to be the synchrotron cooling time for $\gamma_{s}$, $T_{s}=T_{cool}(\gamma_{s})=1/(c_{0}B^{2}\gamma_{s})$ and with such choice of $T_{s}$ the transport equation further simplifies to,
\begin{eqnarray}
    \frac{\partial f}{\partial T} +\frac{\partial}{\partial \Gamma}\left\{\frac{2D_{0}\gamma_{s}^{-\frac{5}{3}}\Gamma^{-\frac{5}{3}}}{\gamma_{s}c_{0}B^2\gamma_{s}}f- \frac{c_{0}B^2(\gamma_{s}\Gamma)^2 f}{\gamma_{s}c_{0}B^2\gamma_{s}}\right\} =
    \frac{\partial}{\partial \Gamma}\left(\frac{D_{0} \gamma_{s}^{-\frac{2}{3}}\Gamma^{-\frac{2}{3}}}{\gamma_{s}^2 c_{0}B^2\gamma_{s}}\frac{\partial f}{\partial \Gamma}\right) - f\frac{\gamma_{s}^{\frac{8}{3}}\alpha \Gamma^{\frac{8}{3}}}{c_{0}B^2\gamma_{s}} + S_{0}\frac{\gamma_{s}^2\Gamma^2}{c_{0}B^2\gamma_{s}},
\end{eqnarray}
which finally takes the following form,
\begin{eqnarray}
    \frac{\partial f}{\partial T} +\frac{\partial}{\partial \Gamma}\left(2a\Gamma^{-\frac{5}{3}}f- \Gamma^2 f \right) = \frac{\partial}{\partial \Gamma}\left(a\Gamma^{-\frac{2}{3}}\frac{\partial f}{\partial \Gamma}\right) - b\Gamma^{\frac{8}{3}}f + S.
\end{eqnarray}
where $a$ and $b$ are the ratios of the synchrotron cooling time to diffusion timescale and escape timescale at $\gamma=\gamma_{s}$, respectively; $S$ is the scaled source term.
They can be defined in the following way,
\begin{eqnarray}\label{eq:coeff_mod}
    a=\frac{D_{0}\gamma_{s}^{-\frac{11}{3}}}{c_{0}B^2},\quad b=\frac{\gamma_{s}^{\frac{5}{3}}\alpha}{c_{0}B^2},\quad S=S_{0}\frac{\gamma_{s}}{c_{0}B^2}\Gamma^2.
\end{eqnarray}

\section{Comparison with Hard-sphere turbulence}

\begin{figure}
    \centering
    \includegraphics[scale=0.68]{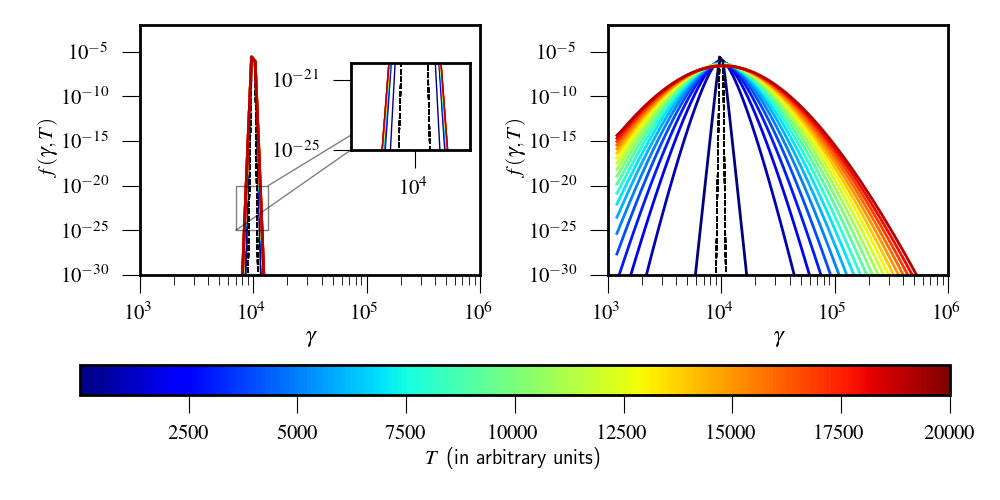}
    \caption{Evolution of an initial Gaussian with mean $10^{4}$ and standard deviation $100$ (shown by a black dashed curve) for two different cases, following Eq.~(\ref{eq:compare}).
    \textbf{Left:} Due to small-scale turbulence ($D=\gamma^{-2/3}$), \textbf{Right:} due to hard-sphere turbulence ($D=\gamma^{2}$).
    The temporal value is depicted by the colorbar.}
    \label{fig:compare}
\end{figure}
In this appendix we show a comparative analysis between the acceleration efficiency of the small-scale turbulence and hard-sphere turbulence.
For this purpose we solve the following Fokker-Planck equation considering different forms for the diffusion coefficient.
\begin{eqnarray}\label{eq:compare}
    \frac{\partial f}{\partial T} +\frac{\partial}{\partial \gamma}\left(\frac{2D}{\gamma}\right)f=\frac{\partial}{\partial \gamma}\left(D\frac{\partial f}{\partial \gamma}\right),
\end{eqnarray}
We consider $D=D_{0}\gamma^{-2/3}$ for the stochastic acceleration due to small-scale turbulence (see section~{\ref{sec:dpp}) and $D=D_{hs}\gamma^{2}$ for the case of hard-sphere turbulence with both $D_{0}=D_{hs}=1$.

In Fig.~\ref{fig:compare} we show the temporal evolution of the initial distribution function for both the case scenarios.  
In the left plot, due to small-scale turbulence, the spread of the initial distribution function increases owing to the acceleration of particles. 
However, the spreading of the distribution function happens slowly compared to the plot shown in the right panel.
Such an evolution is due to the larger acceleration time ($\tau_{acc}\sim \gamma^2/D\propto\gamma^{8/3}$) for the small-scale turbulence as compared to the right one where $\tau_{acc}$ is constant. 
This exercise clearly shows that the acceleration due to the small-scale turbulence is less efficient as compared to the hard-sphere case, however one should keep in mind that the result is largely dependent on the choice of $D_{0}$ and $D_{hs}$ (see Fig.~\ref{fig:acceleration}).
The purpose of this analysis is to compare the evolution of the distribution function for two mathematically different form of $D$.

\section{Evolution of the distribution function with different escape timescale }
In this appendix we show additional figures (figs.~\ref{fig:1e-4_escape} and \ref{fig:1e-5_escape}) for the evolution of the distribution function by solving Eq.~(\ref{eq:transport}) with $b=10^{-4}$ and $10^{-5}$.
The evolution is computed for different values of $a$ and $S=0$. 

\begin{figure}
    \centering
    \includegraphics[scale=0.5]{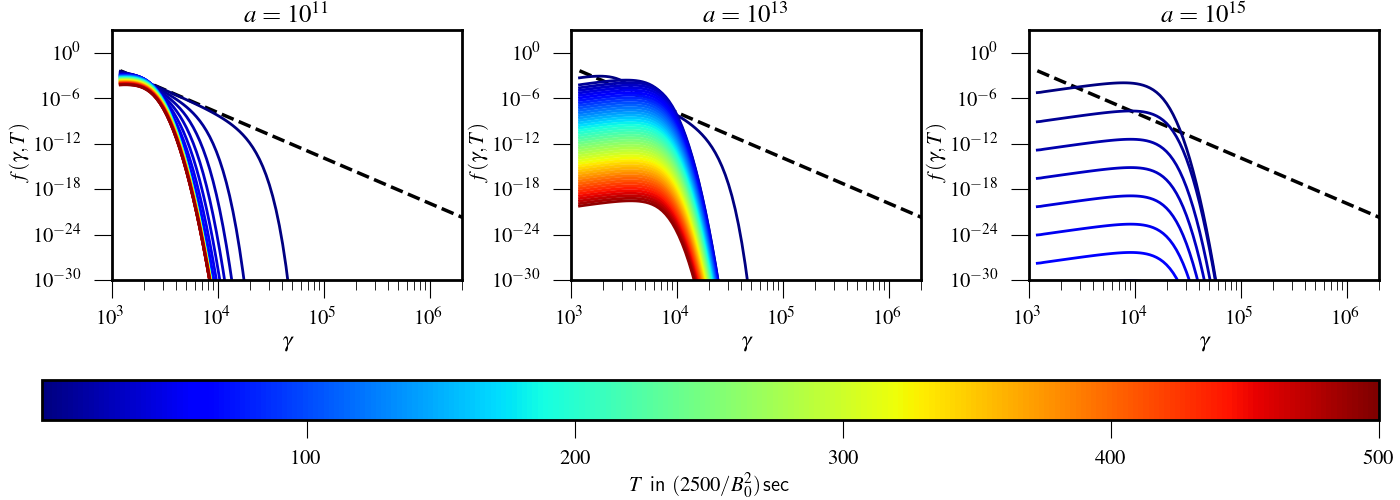}
    \caption{Evolution of an initial power-law energy distribution of the form $\gamma^{-6}$ following Eq.~(\ref{eq:transport}) considering synchrotron loss process with different values for $a$ and $b=10^{-5}$.
    The values for $S$ is considered as zero.
    The initial distribution is shown with the black dashed curve.
    {Different color of the distribution function corresponds to different time of evolution, as illustrated by the colorbar. 
    To account for the varying magnetic field values observed in different astrophysical systems and the resulting variation in temporal units, the unit time is specified in terms of a variable magnetic field.}}
    \label{fig:1e-5_escape}
\end{figure}

\begin{figure}
    \centering
    \includegraphics[scale=0.5]{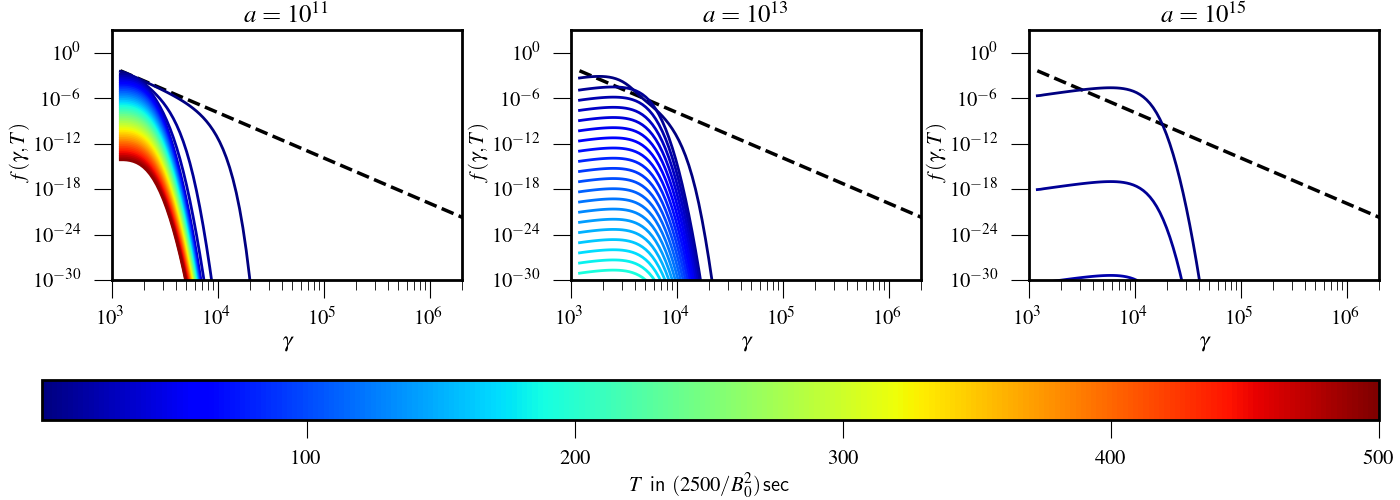}
    \caption{Evolution of an initial power-law energy distribution of the form $\gamma^{-6}$ following Eq.~(\ref{eq:transport}) considering synchrotron loss process with different values for $a$ and $b=10^{-4}$.
    The values for $S$ is considered as zero.
    The initial distribution is shown with the black dashed curve.
    {Different color of the distribution function corresponds to different time of evolution, as illustrated by the colorbar. 
    To account for the varying magnetic field values observed in different astrophysical systems and the resulting variation in temporal units, the unit time is specified in terms of a variable magnetic field.}}
    \label{fig:1e-4_escape}
\end{figure}

\section{Computation of transport coefficients for small-scale anisotropic Alfv\`en wave turbulence spectrum}
\label{sec:deriv_anisotropy}
For computing the transport coefficients for small-scale anisotropic turbulence, we consider the form of the turbulence spectrum such that it can mimic the behaviour of realistic Alfv\`enic turbulence upto a certain degree \citep{yan_2002}.
In particular, we consider the turbulence spectrum of the following form,
\begin{eqnarray}
    Y_{ij}(k)=P_{aniso}\left(\delta_{ij}-\frac{k_{i}k_{j}}{k_{\perp}^2}\right)\Theta\left(k_{\perp} - m'k_{g}\right)\delta\left(\frac{k_{||}}{m''k_{g}}-1\right)k_{\perp}^{-\alpha},
\end{eqnarray}
where $\Theta$ corresponds to Heaviside theta function and $P_{aniso}$ being the injected turbulent power; $k_{\perp}$ and $k_{||}$ are the perpendicular and parallel wave vector components; $m'k_{g}$ and $m''k_{g}$ are the respective scales where power corresponding to $k_{\perp}$ and $k_{||}$ are injected. 
Note the difference between the above spectrum with the isotropic one given by Eq.~(\ref{eq:turb_spec}), the isotropic part here corresponds to the isotropy in the plane perpendicular to $k_{||}$ and unlike the earlier one the injection of energy is happening at different scales for $k_{\perp}$ and $k_{||}$ separately, which is governed by the value of $m'$ and $m''$ respectively.
Moreover the above spectrum allows for the energy to cascade along $k_{\perp}$ direction, while a single scale injection is considered along $k_{||}$.
With such a spectrum the equipartition of energy implies the form of $P_{aniso}$ as the following,
\begin{eqnarray}
    P_{aniso} \simeq\frac{\alpha-2}{2\pi m'k_{g}(m''k_{g})^{2-\alpha}}.
\end{eqnarray}
The positivity constraint of the power implies $\alpha>2$.
Such a turbulence spectrum takes the following form in the polarisation space,
\begin{itemize}
    \item $\DS R_{\mathcal{RR}} = \frac{1}{2} \frac{B_{0}^{2} V_{A}^{2}}{c^{2}}P_{aniso}\Theta\left(k_{\perp} - m'k_{g}\right)\delta\left(\frac{k_{||}}{m''k_{g}}-1\right)k_{\perp}^{-\alpha}$;
    \item $\DS R_{\mathcal{LL}} = \frac{1}{2} \frac{B_{0}^{2} V_{A}^{2}}{c^{2}}P_{aniso}\Theta\left(k_{\perp} - m'k_{g}\right)\delta\left(\frac{k_{||}}{m''k_{g}}-1\right)k_{\perp}^{-\alpha}$;
    \item $\DS R_{\mathcal{LR}} = \frac{1}{2} \frac{B_{0}^{2} V_{A}^{2}}{c^{2}}P_{aniso}\Theta\left(k_{\perp} - m'k_{g}\right)\delta\left(\frac{k_{||}}{m''k_{g}}-1\right)k_{\perp}^{-\alpha}e^{2\iota\psi}$;
    \item $\DS R_{\mathcal{RL}}= \frac{1}{2} \frac{B_{0}^{2} V_{A}^{2}}{c^{2}}P_{aniso}\Theta\left(k_{\perp} - m'k_{g}\right)\delta\left(\frac{k_{||}}{m''k_{g}}-1\right)k_{\perp}^{-\alpha}e^{-2\iota\psi}$;
\end{itemize}
With such turbulent spectrum, $D_{pp}$ takes the following form,
\begin{eqnarray}\label{eq:Dpp_power_law}
    D_{pp}= \frac{\Omega\left(1-\mu^2\right)}{4}2\pi^{2}m^2c^2\frac{V_{A}^{2}}{c^{2}}m''k_{g}P_{aniso}\mathcal{R}e\left[\int_{m'k_{g}}^{\infty}k_{\perp}^{-\alpha+1}d k_{\perp}\left\{J_{\frac{\omega}{\Omega}-\frac{m''k_{g}v_{||}}{\Omega}+1}^2\left(\frac{k_{\perp}v_{\perp}}{\Omega}\right)+J_{\frac{\omega}{\Omega}-\frac{m''k_{g}v_{||}}{\Omega}-1}^2\left(\frac{k_{\perp}v_{\perp}}{\Omega}\right)\right .\right .
    \\ \nonumber
    \left.\left. +2J_{\frac{\omega}{\Omega}-\frac{m''k_{g}v_{||}}{\Omega}+1}\left(\frac{k_{\perp}v_{\perp}}{\Omega}\right)J_{\frac{\omega}{\Omega}-\frac{m''k_{g}v_{||}}{\Omega}-1}\left(\frac{k_{\perp}v_{\perp}}{\Omega}\right) \right\} \right]
\end{eqnarray}
Note that the integral over $k_{\perp}$ resembles a variation of the Weber-Schafheitlin type integral and the integration is performed by bounding the upper limit of the the integral due to the constraint given in Eq.~(\ref{eq:aniso_const}).
An analytical evaluation of the integration over $k_{\perp}$ is possible and takes the following form\footnote{Mathematica code: 
    \includegraphics[scale=0.51]{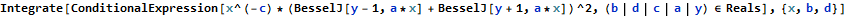}},
\begin{equation}
\begin{aligned}
\int_{b}^{d}x^{-c} \left(J^{2}_{y-1}(ax)+J^{2}_{y+1}(ax)+2J_{y+1}(ax)J_{y-1}(ax)\right) \,dx
\\ 
=\frac{2^{-2 (y+1)} (b d)^{-c}}{\Gamma (y)^2} \left(\frac{a^{2 y+2}}{y^2 (y+1)^2 (-c+2 y+3)} \left(b^c d^{2 y+3} \, _2F_3\left(y+\frac{3}{2},-\frac{c}{2}+y+\frac{3}{2};y+2,-\frac{c}{2}+y+\frac{5}{2},2 y+3;-a^2
   d^2\right)
\right.\right.
\\
\left.\left.
-b^{2 y+3} d^c \, _2F_3\left(y+\frac{3}{2},-\frac{c}{2}+y+\frac{3}{2};y+2,-\frac{c}{2}+y+\frac{5}{2},2 y+3;-a^2 b^2\right)\right)
\right.
\\
\left.
+\frac{16}{a^2 b d (c-2 y+1)}
   \left(d^{c+1} (a b)^{2 y} \, _2F_3\left(y-\frac{1}{2},-\frac{c}{2}+y-\frac{1}{2};y,-\frac{c}{2}+y+\frac{1}{2},2 y-1;-a^2 b^2\right)
\right.\right.
\\
\left.\left.
-b^{c+1} (a d)^{2 y} \,
   _2F_3\left(y-\frac{1}{2},-\frac{c}{2}+y-\frac{1}{2};y,-\frac{c}{2}+y+\frac{1}{2},2 y-1;-a^2 d^2\right)\right)\right.
\\
\left.
+\frac{8 \Gamma (y)}{(c+1) (-c+2 y+1) \Gamma (y+2)} \left(d b^c (c-2 y-1) (a
   d)^{2 y} \, _2F_3\left(y+\frac{1}{2},y+1;y,y+2,2 y+1;-a^2 d^2\right)\right.\right.
\\
\left.\left.
+b d^c (-c+2 y+1) (a b)^{2 y} \, _2F_3\left(y+\frac{1}{2},y+1;y,y+2,2 y+1;-a^2 b^2\right)
\right.\right.
\\
\left.\left.
-2 (y+1) \left(b
   d^c (a b)^{2 y} \, _2F_3\left(y+\frac{1}{2},-\frac{c}{2}+y+\frac{1}{2};y,-\frac{c}{2}+y+\frac{3}{2},2 y+1;-a^2 b^2\right)\right.\right.\right.
\\
\left.\left.\left.
-d b^c (a d)^{2 y} \,
   _2F_3\left(y+\frac{1}{2},-\frac{c}{2}+y+\frac{1}{2};y,-\frac{c}{2}+y+\frac{3}{2},2 y+1;-a^2 d^2\right)\right)\right)\right)
\end{aligned}
\end{equation}
However due to the presence of the combinations of Gamma functions $\Gamma(...)$ and Confluent hypergeometric fuctions $_{2}F_{3}(...)$ the result of the integration becomes very sensitive on the values of different parameters, $a,\,b,\,c\,\text{and}\,d$.
Owing to such reason we perform the integration numerically obeying the constraint given by Eq.~(\ref{eq:aniso_const}). 

\section{Transport coefficient for fast magnetosonic wave}\label{sec:fast}
We show the calculation of the momentum transport coefficient for the scenario when the small-scale turbulence is mediated via compressional fast waves. 
For simplicity we consider the dispersion relation of the fast wave in a cold plasma medium, which takes the following form \citep[see Eq.~13.3.1 in][]{schlickeizer_2002},
\begin{eqnarray}
  \omega=V_{A}k  
\end{eqnarray}
where $V_{A}$ is the Alfv\`{e}n velocity and $k=|\vec{k}|$.
With such dispersion relation Eq.~(\ref{eq:resonance_mod}) becomes the following, 
\begin{eqnarray*}
\frac{\gamma m' k_{g} V_{A}}{\Omega_{NR}} - \frac{m' k_{g} c \sqrt{1-\frac{1}{\gamma^2}}\mu x}{\Omega_{NR}}\gamma - \frac{m' k_{g}c \sqrt{1-\frac{1}{\gamma^2}}}{\Omega_{NR}}\gamma \sqrt{1-x^2}\sqrt{1-\mu^2} = Q,
\end{eqnarray*}
With such a condition, we compute the region of validity for $x$ from the following equation, 
\begin{eqnarray}
\label{eq:fast}
Q_{min}\leq\frac{\gamma m' k_{g} V_{A}}{\Omega_{NR}} - \frac{m' k_{g} c \sqrt{1-\frac{1}{\gamma^2}}\mu x}{\Omega_{NR}}\gamma - \frac{m' k_{g}c \sqrt{1-\frac{1}{\gamma^2}}}{\Omega_{NR}}\gamma \sqrt{1-x^2}\sqrt{1-\mu^2} \leq Q_{max},
\end{eqnarray}
for fast waves and thereby calculate the momentum diffusion coefficient following Eq.~(\ref{eq:Dppfinal}).

Below we investigate the impact of various parameter values on the pitch-angle-averaged momentum diffusion coefficient for isotropic fast wave turbulence.
The results are presented in Fig.~\ref{fig:fast}, which comprises of four panels displaying the impact of different parameters on the diffusion coefficient.
The diffusion coefficient can be observed to follow a power-law like trend with an index of $-2/3$ with the particle Lorentz factor $\gamma$ in all of the panels, similar to that of the Alfv\`{e}nic turbulence as shown in section~\ref{sec:dpp}.

Panel (a) of the figure investigates the influence of different values of the mean magnetic field $B$ on the diffusion coefficient. 
It can be observed that the diffusion coefficient increases with the magnetic field strength.
In panel (b), the effect of the parameter $m'$ on the diffusion coefficient is examined. 
It is observed that a smaller energy injection scale, corresponds to larger values of $m'$, results in a reduced influence of turbulence on the charged particles, causing the diffusion coefficient to decrease.
Panel (c) investigates the effect of the Alfv\'{e}n velocity on the momentum diffusion coefficient. 
The results indicate that a decrease in the Alfv\'{e}n velocity of the underlying fluctuations leads to a reduction in momentum diffusion.
In addition, panel (d) examines the effect of the parameter $\sigma$ on the momentum diffusion coefficient. 
The results indicate that the momentum diffusion coefficient increases as $\sigma$ increases, due to particles interacting with an increasing number of waves as $\sigma$ increases.

Moreover, comparing the value of the diffusion coefficients by modulating various parameters with the one observed for the Alfv\`{e}n waves (as discussed in section~\ref{sec:dpp}, see also Fig.~\ref{fig:diff_gamm}), we find the values to be of the same order which ultimately resonate with the fact that the nature of the turbulence becomes degenerate to the non-thermal particles whose gyro-radius is higher that the turbulence correlation length.  

\begin{figure}
    \centering
     \includegraphics[scale=0.31]{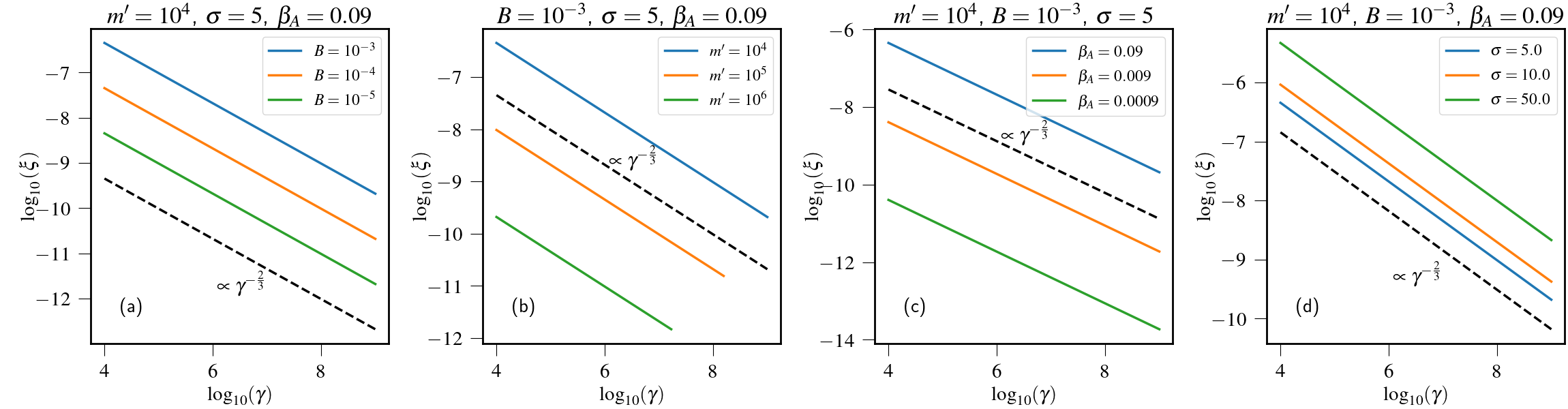}
    \caption{{Figure demonstrating the dependence of the pitch-angle-averaged momentum diffusion coefficient ($\xi$) on particle Lorentz factor $\gamma$ considering an isotropic single-scale turbulence injection spectrum for fast magnetosonic waves in cold plasma.}}
    \label{fig:fast}
\end{figure}

% Don't change these lines
\bsp	% typesetting comment
\label{lastpage}
\end{document}